\documentclass[fp,twocolumn]{jpsj3}
\usepackage{txfonts,bm,amsmath,amssymb,url,float,cases,braket,xcolor,graphicx}

\begin{document}

\title{
Analysis of Magnetoacoustic Quadrupole Resonance and Application to Probe Quadrupole Degrees of Freedom in Quantum Magnets
}

\author{
Masashige Matsumoto$^1$\thanks{E-mail address: matsumoto.masashige@shizuoka.ac.jp} and Mikito Koga$^2$
}

\inst{
$^1$Department of Physics, Faculty of Science, Shizuoka University, Shizuoka 422-8529, Japan \\
$^2$Department of Physics, Faculty of Education, Shizuoka University, Shizuoka 422-8529, Japan
}

\recdate{March 27, 2020}

\abst{
Motivated by the recent progress of high-frequency ultrasonic measurements,
we propose a theory of magnetoacoustic resonance as a microscopic probe for quadrupole degrees of freedom
hidden in magnetic materials.
A local strain driven by an acoustic wave couples to electronic states of a magnetic ion
through various quadrupole--strain couplings, and this provides a periodically time-dependent oscillating field.
As a typical two-level system with the quadrupole, we consider a non-Kramers doublet
and investigate single- and multiphonon-mediated transition processes on the basis of the Floquet theory.
An analytic form of the transition probability is derived within the weak coupling theory,
which helps us analyze the magnetoacoustic quadrupole resonance.
We apply the theory to realistic non-Kramers doublet systems for the $f^2$ configuration in $O_h$ and $D_{4h}$ symmetries,
and discuss how to identify the relevant quadrupole by controlling the quadrupole--strain coupling
with an applied magnetic field in ultrasonic measurements.
}

\maketitle


\renewcommand{\H}{{\mathcal H}}
\newcommand{\Sch}{Schr\"{o}dinger}

\newcommand{\bH}{{\bm H}}
\newcommand{\bh}{{\bm h}}
\newcommand{\bJ}{{\bm J}}
\newcommand{\bQ}{{\bm Q}}
\newcommand{\bq}{{\bm q}}
\newcommand{\bk}{{\bm k}}
\newcommand{\br}{{\bm r}}
\newcommand{\bS}{{\bm S}}
\newcommand{\be}{{\bm e}}


\section{Introduction}

In conventional magnets, high-rank multipoles such as quadrupoles
are usually hidden by active magnetic dipoles and magnetic ordering.
Magnetic properties can be probed by magnetic susceptibility, nuclear magnetic resonance (NMR),
electron spin resonance (ESR),
\cite{Slichter-1990}
 and neutron scattering measurements.
\cite{Squires-2012}
Recently, much attention has been paid to some magnetic materials in which magnetic dipoles become inactive
owing to geometric frustration, competition with the Kondo effect, and stabilizing nonmagnetic crystal-field states.
In this case, high-rank multipoles play an important role than the magnetic dipoles.
\cite{Santini-2009, Kuramoto-2009}
Among them, the quadrupole degrees of freedom in $f$-electron materials such as Ce- and Pr-based compounds
have been actively investigated.
\cite{Thalmeier-2019, Onimaru-2016}

The conventional experimental techniques for directly probing quadrupoles and quadrupole ordering
are measurements of the softening of elastic constants and resonant X-ray scattering.
There is still room for the development of more precise and usable measurements of quadrupoles
with the aid of the recent progress in the application of high-frequency ultrasonic waves to spintronics.
\cite{Sasaki-2019,Puebla-2020}
Our work is also motivated by an epoch-making ultrasonic measurement
that demonstrated its power as a tool for evaluating the vacancy concentration
in the surface layer of a silicon wafer with high sensitivity.
\cite{Goto-2006,Mitsumoto-2014}
The high sensitivity can be explained by a strongly enhanced quadrupole--strain coupling
originating from a vacancy orbital state, although this has not been confirmed yet.

Dynamical measurements based on resonance phenomena
provide important information on physical and chemical properties of materials.
By analogy with NMR and ESR, we have recently suggested a new type of photon-assisted magnetoacoustic measurement
with various quadrupole couplings in magnetic multiplets,
\cite{Koga-2020}
which is related to the local charge distribution modified by a lattice deformation.
The quadrupole--strain couplings can be driven by an acoustic wave
propagating in the lattice or on the surface layer.
This allows a resonance measurement mediated by quadrupole components of local electronic states.
The suggested photon-assisted magnetoacoustic resonance can also be applied
to optical control in quantum spin devices.
\cite{Koga-2020,Okazaki-2018,Chen-2018,Udvarhelyi-2018}
In this paper, we extend our previous study of magnetoacoustic resonance
\cite{Koga-2020}
to various doublet states with quadrupole couplings using the Floquet theory originally formulated by Shirley
for a two-level system coupled to a periodically time-dependent oscillating field.
\cite{Shirley-1965}
This theory covers the strong coupling region as well as the weak coupling limit,
and is useful for describing the fundamental properties of magnetoacoustic quadrupole resonance.

An earlier study of acoustic quadrupole measurement was performed
with the nuclear quadrupole resonance in NaCl.
\cite{Proctor-1956-2}
As an example of ESR, a Cr$^{3+}$ ion in Al$_2$O$_3$ (ruby) was studied
by ultrasonic measurement, and a quadrupole transition was reported.
\cite{Tucker-1961}
As a recent optical probe, a quadrupole transition in Sr$_2$CoGe$_2$O$_7$ under a high magnetic field was reported.
\cite{Akaki-2017}
In this case, there is no space inversion symmetry at the Co$^{2+}$ ion site,
and this enables the Co$^{2+}$ spin to couple to the electric field component of light through a quadrupole.
\cite{Akaki-2017,Hou-1965,Mims-1976,Arima-2007,Matsumoto-2017}
Thus, such quadrupole resonances in quantum magnets already led to some achievements with the potential for more developments.
Comparing the acoustic and optical quadrupole resonances mentioned above,
the former has an advantage since the quadrupole--strain coupling can exist
even in the presence of inversion symmetry.
For metallic samples, an acoustic wave can penetrate deeper into the bulk than an optical wave,
and this is another advantage of the acoustic quadrupole resonance.

For our purpose, a non-Kramers doublet is the most appropriate candidate,
which can be realized as an atomic ground state with an integer spin or angular momentum
for an even $d$- or $f$-electron configuration.
It is important that quadrupole--strain couplings can be controlled by rotating a magnetic field,
since the degeneracy of a non-Kramers doublet is lifted by a Van Vleck process through excited states.
We also reveal unknown properties of the quadrupole dynamics driven by an acoustic wave.
In fact, rich quadrupole physics is expected in non-Kramers doublet systems,
such as Pr-based compounds with a well-separated doublet ground state from other crystal-field excited states.
\cite{Onimaru-2005, Nakanishi-2018, Taniguchi-2019,Yanagisawa-2019}

This paper is organized as follows.
In Sect. \ref{sec:s=1}, we focus on an $S=1$ system with an easy-axis anisotropy
and investigate details of the transition probability with an analytic formula derived from the weak coupling theory.
The theory is applied to realistic non-Kramers doublet systems
in cubic $O_h$ and tetragonal $D_{4h}$ crystal fields in Sects. \ref{sec:oh} and \ref{sec:d4h}, respectively.
Its application to probe the quadrupole order is also discussed in Sect. \ref{sec:order}.
The last section gives a summary and discussion.
In Appendix\ref{appendix:matrix}, spin and quadrupole operators for an $S=1$ system are given in matrix forms.
In Appendix\ref{appendix:matter-tensor},
the quadrupole--strain coupling and the magnetic-field-dependent Hamiltonian for the non-Kramers doublet
are derived from the fourth-rank matter tensor.
Appendix\ref{appendix:weak} gives details of the weak coupling theory for the transition probability.

\section{Basic Formulation}
\label{sec:s=1}

To demonstrate the magnetoacoustic quadrupole resonance in quantum magnets,
we first focus on a tetragonal system of $D_{4h}$ symmetry.
This system is easy to handle and provides a fundamental model for understanding the basic properties of the quadrupole resonance.

\subsection{Effective Hamiltonian for $S=1$ in $D_{4h}$ symmetry}
\label{sec:s=1-d4h}

Let us begin with the following local Hamiltonian of an $S=1$ quantum spin under a finite magnetic field $\bH$:
\begin{align}
\H = \H_0 + \H',
\label{eqn:H0}
\end{align}
with
\begin{align}
\H_0 = - D S_z^2,~~~~~~
\H' = - \bh\cdot\bS.
\label{eqn:H0-2}
\end{align}
Here, $\bS=(S_x,S_y,S_z)$ is the spin operator.
$\bh = g\mu_{\rm B} \bH$, where $g$ is the $g$-factor and $\mu_{\rm B}$ is the Bohr magneton.
$D(>0)$ is a constant representing a uniaxial anisotropy,
where a doublet ($S_z=\pm 1$) and singlet ($S_z=0$) energy level scheme is realized for $H=0$.
For $S=1$, note that the spin Hamiltonian $\H_0$ in Eq. (\ref{eqn:H0-2}) is isotropic around the $z$-axis even in $D_{4h}$ symmetry.
In general, there are quadrupole degrees of freedom in an $S\ge 1$ spin.
For $S=1$, matrix forms of the spin and quadrupole operators are given by Eqs. (\ref{eqn:S-mat}) and (\ref{eqn:O-mat}), respectively.

When the magnetic field is applied in the $xy$-plane [$\bH=(H_x,H_y,0)$],
the energy of the doublet does not split linearly with the field, but it splits quadratically with the field.
In this sense, the $S_z=\pm 1$ doublet can be regarded as a non-Kramers doublet.
Then, we treat $\H'$ as a perturbation
and derive an effective Hamiltonian for the non-Kramers doublet under the field in the $xy$-plane.
Note that $\H'$ has no matrix element among the doublet states [see $S_x$ and $S_y$ in Eq. (\ref{eqn:S-mat})].

Let us represent $\ket{\alpha}$ and $\ket{\beta}$ as the non-Kramers doublet states of the unperturbed Hamiltonian $\H_0$,
whereas $\ket{\gamma}$ represents other energy eigenstates of $\H_0$.
Their energy eigenvalues are expressed as $E_\alpha^{(0)}=E_\beta^{(0)}\neq E_\gamma^{(0)}$, respectively.
On the basis of the perturbation theory, the effective Hamiltonian is given by the following form for $\bH=(H_x,H_y,0)$:
\begin{align}
&\braket{\alpha|\H_{\rm eff}|\beta}
= - \sum_\gamma \frac{ \braket{\alpha|\H'|\gamma} \braket{\gamma|\H'|\beta} }{E_\gamma^{(0)} - E_\alpha^{(0)}} \cr
&= - \sum_\gamma \frac{1}{E_\gamma^{(0)} - E_\alpha^{(0)}} \cr
&~~~\times
\Bigl[
    h_x^2 \braket{\alpha|S_x|\gamma} \braket{\gamma|S_x|\beta}
+ h_y^2 \braket{\alpha|S_y|\gamma} \braket{\gamma|S_y|\beta} \cr
&~~~~~~
+ h_x h_y \left( \braket{\alpha|S_x|\gamma} \braket{\gamma|S_y|\beta}
                         + \braket{\alpha|S_y|\gamma} \braket{\gamma|S_x|\beta} \right)
\Bigr] \cr
&= - \sum_\gamma \frac{1}{2\left(E_\gamma^{(0)} - E_\alpha^{(0)}\right)} \cr
&~~~\times
\Bigl[
(h_x^2+h_y^2) \left(\braket{\alpha|S_x|\gamma} \braket{\gamma|S_x|\beta}+\braket{\alpha|S_y|\gamma} \braket{\gamma|S_y|\beta} \right) \cr
&~~~
+ (h_x^2-h_y^2) \left(\braket{\alpha|S_x|\gamma} \braket{\gamma|S_x|\beta}-\braket{\alpha|S_y|\gamma} \braket{\gamma|S_y|\beta} \right) \cr
&~~~
+ 2h_x h_y \left( \braket{\alpha|S_x|\gamma} \braket{\gamma|S_y|\beta}
                         + \braket{\alpha|S_y|\gamma} \braket{\gamma|S_x|\beta} \right)
\Bigr].
\label{eqn:H-eff}
\end{align}
This shows that the non-Kramers doublet couples to the magnetic field
in a quadratic form through quadrupole degrees of freedom.
$h_x^2+h_y^2$ couples to the $S_x^2+S_y^2$ type quadrupole,
while $h_x^2-h_y^2$ and $2h_x h_y$ couple to the $O_v=S_x^2-S_y^2$ and $O_{xy}=S_x S_y + S_y S_x$ type quadrupoles, respectively.

For the Hamiltonian in Eq. (\ref{eqn:H0}), $E_\alpha^{(0)}=E_\beta^{(0)}=-D$ and $E_\gamma^{(0)}=0$.
Within the doublet states, $\H_{\rm eff}$ is expressed in the following $2\times 2$ matrix form:
\begin{align}
\H_{\rm eff}
&= - \frac{1}{2D}
\left[
(h_x^2+h_y^2) \bm{1}
+ (h_x^2-h_y^2) O_v
+ 2h_x h_y O_{xy}
\right] \cr
&= - \frac{h^2}{2D}
\left[
\begin{pmatrix}
1 & 0 \cr
0 & 1
\end{pmatrix}
+
\begin{pmatrix}
0 & e^{-i2\theta} \cr
e^{i2\theta} & 0
\end{pmatrix}
\right].
\label{eqn:H-eff-1}
\end{align}
Here, $\bm{1}$ represents the unit matrix.
For $S=1$, the coefficients between the magnetic field and ($O_v$, $O_{xy}$) quadrupoles become the same.
Equation (\ref{eqn:H-eff-1}) is covered by Eq. (\ref{eqn:H-h-d4h}) for $\bh=(h_x,h_y,0)$,
where the latter equation is the general form of the field dependence of the Hamiltonian in $D_{4h}$ symmetry.
In Eq. (\ref{eqn:H-eff-1}), the magnetic field is represented by $\bh=(h_x,h_y,0)=h(\cos\theta,\sin\theta,0)$,
where $\theta$ is the angle of the field measured from the $x$-axis.
$O_v$ and $O_{xy}$ are quadrupole operators whose matrix forms are given by [see Eq. (\ref{eqn:O-mat})]
\begin{align}
O_v =
\begin{pmatrix}
0 & 1 \cr
1 & 0
\end{pmatrix},~~~~~~
O_{xy} =
\begin{pmatrix}
0 & -i \cr
i & 0
\end{pmatrix}.
\end{align}
The first term in Eq. (\ref{eqn:H-eff-1}) is already diagonal.
Since it represents a uniform energy shift for the doublet, we omit it in the following discussion.
The effective Hamiltonian can be diagonalized by the following unitary transformation:
\begin{align}
\tilde{\H}_{\rm eff} =
U^\dagger \H_{\rm eff} U = - \frac{h^2}{D} \frac{1}{2}
\begin{pmatrix}
1 & 0 \cr
0 & -1
\end{pmatrix}.
\label{eqn:H-eff-2}
\end{align}
Here, we introduced
\begin{align}
U = \frac{1}{\sqrt{2}}
\begin{pmatrix}
ie^{-i\theta} & -e^{-i\theta} \cr
ie^{i\theta} & e^{i\theta}
\end{pmatrix}.
\end{align}
In Eq. (\ref{eqn:H-eff-2}), $h^2/D$ represents the energy splitting of the non-Kramers doublet under the field,
which is termed as the Van Vleck splitting.
In the diagonalized basis of $\tilde{\H}_{\rm eff}$, the matrix forms of the quadrupole operators are transformed as
\begin{align}
&\tilde{O}_v = U^\dagger O_v U =
\begin{pmatrix}
\cos{2\theta} & \sin{2\theta} \cr
\sin{2\theta} & -\cos{2\theta}
\end{pmatrix},
\label{eqn:O-2} \\
&\tilde{O}_{xy} = U^\dagger O_{xy} U =
\begin{pmatrix}
\sin{2\theta} & -\cos{2\theta} \cr
-\cos{2\theta} & -\sin{2\theta}
\end{pmatrix}.
\nonumber
\end{align}

\subsection{Floquet theory}
\label{sec:floquet}

The local strain driven by an acoustic wave is classified by the point group symmetry.
It couples to a spin through a quadrupole when the local strain and quadrupole belong to the same irreducible representation.
In the $D_{4h}$ symmetry, the general form of the quadrupole--strain coupling is given by Eq. (\ref{eqn:H-d4h-2}),
where the $O_v$ quadrupole couples to the $\varepsilon_{xx}-\varepsilon_{yy}$ strain.
This is because they belong to the $\Gamma_3$ ($B_{1g}$) representation.
Here, $\varepsilon_{ij}$ denotes the strain tensor defined by Eq. (\ref{eqn:strain}).
Similarly, the $O_{xy}$ quadrupole couples to the $2\varepsilon_{xy}$ strain,
where they belong to the $\Gamma_4$ ($B_{2g}$) representation.
In this subsection, we focus on the $\varepsilon_{xx}-\varepsilon_{yy}$ strain with a periodic oscillation
coupled to the $O_v=S_x^2 - S_y^2$ quadrupole,
since the roles of the $O_v$ and $O_{xy}$ quadrupoles are interchanged by the $\theta\rightarrow \theta - \pi/4$ transformation.

The effective Hamiltonian for the non-Kramers doublet is given by Eq. (\ref{eqn:H-eff-2}) in the diagonal form,
and the transformed quadrupole $\tilde{O}_v$ in Eq. (\ref{eqn:O-2}) is used for the periodic vibration of the local strain.
The time-dependent effective Hamiltonian is then expressed in the following form:
\begin{align}
&\H_{\rm eff}(t) = \H_{\varepsilon_0} + \H_A(t,\theta), \cr
&\H_{\varepsilon_0} = - \frac{\varepsilon_0}{2}
\begin{pmatrix}
1 & 0 \cr
0 & -1
\end{pmatrix}, \cr
&\H_A(t,\theta) =
\frac{1}{2}
\begin{pmatrix}
A_L(\theta) & A_T(\theta) \cr
A_T(\theta) & -A_L(\theta)
\end{pmatrix}
\cos{\omega t},
\label{eqn:H-mat}
\end{align}
with
\begin{align}
A_L(\theta)=A\cos{2\theta},~~~~~~
A_T(\theta)=A\sin{2\theta}.
\label{eqn:A}
\end{align}
Here, $\varepsilon_0 = h^2/D = (g\mu_{\rm B}H)^2/D$ is the Van Vleck splitting.
$A_L(\theta)$ and $A_T(\theta)$ represent the longitudinal (diagonal) and transverse (off-diagonal) components
of the local vibration, respectively.
$\omega$ is the angular frequency of the periodic vibration
and $A$ represents the coupling constant between the $O_v$ quadrupole and $\varepsilon_{xx}-\varepsilon_{yy}$ strain.
$\theta$ is the angle of the magnetic field measured from the $x$-axis.
The $A_L(\theta)$ term modifies the excitation gap periodically,
whereas the $A_T(\theta)$ term gives rise to a transition between the two states.

The effective Hamiltonian $\H_{\rm eff}(t)$ in Eq. (\ref{eqn:H-mat}) is valid for $g\mu_{\rm B}H \ll D$.
It describes the time evolution of the two-level system.
As pointed out in Ref. \ref{ref:Koga-2020},
the amplitudes of the longitudinal [$A_L(\theta)$] and transverse [$A_T(\theta)$] components
change with the rotation of the magnetic field,
reflecting the fact that the wave functions of the doublet are modified by the field direction.
This appears as the interchange between $A_L(\theta)$ and $A_T(\theta)$,
where both of which are the keys in determining the character of the resonance.
Thus, the magnetoacoustic quadrupole resonance can be controlled by tuning the direction of the magnetic field.

We remark that the Hamiltonian in Eq. (\ref{eqn:H-mat}) is essentially the same as that for an $S=1/2$ spin
under a static magnetic field along the $z$-axis with a periodically oscillating field tilted from the $z$-axis in the $zx$-plane.
This model contains both the longitudinal and transverse components,
and was studied theoretically and experimentally to find multiphoton resonances in pulse electron paramagnetic resonance.
\cite{Gromov-2000}
The present study can be then considered as an extension of this conventional photonic measurement
to the magnetoacoustic quadrupole resonance for the transition probability.

On the basis of the formulation by Shirley, the time-dependent \Sch~equation
reduces to the eigenvalue problem of the Floquet Hamiltonian $\H_F$.
The matrix elements of $\H_F$ are expressed as
\cite{Shirley-1965}
\begin{align}
\braket{\alpha n | \H_F | \beta m}
&= \H_{\varepsilon_0}^{\alpha\beta} \delta_{n=m} + n\omega \delta_{\alpha=\beta} \delta_{n=m} \cr
&~~~
+ \H_A^{\alpha\beta}(\theta) \left( \delta_{n-m=1} + \delta_{n-m=-1} \right).
\label{eqn:HF}
\end{align}
Here, the index $n~(=0,\pm 1, \pm 2, \cdots)$ of the Floquet state $\ket{\alpha n}$
($\alpha=E_1, E_2$) ($E_1=-\varepsilon_0/2, E_2=\varepsilon_0/2$)
corresponds to the $e^{i n\omega t}$ time-dependent wave function.
\cite{Shirley-1965}
In Eq. (\ref{eqn:HF}), $\delta_{n=m}$ and similar symbols represent the Kronecker delta.
$\H_{\varepsilon_0}^{\alpha\beta}$ and $\H_A^{\alpha\beta}(\theta)$ represent the matrix elements of
$\H_{\varepsilon_0}$ and $\H_A(\theta)$ in Eq. (\ref{eqn:H-mat}), respectively.
The explicit matrix form of $\H_F$ is given by
\cite{Shirley-1965}
\begin{align}
\begin{footnotesize}
\H_F =
\begin{pmatrix}
\cdot & \cdot & \cdot & \cdot & \cdot & \cdot & \cdot & \cdot \cr
\cdot & -\frac{\varepsilon_0}{2}-\omega & 0 & a_L & a_T & 0 & 0 & \cdot \cr
\cdot & 0 & \frac{\varepsilon_0}{2}-\omega & a_T & -a_L & 0 & 0 & \cdot \cr
\cdot & a_L & a_T & -\frac{\varepsilon_0}{2} & 0 & a_L & a_T & \cdot \cr
\cdot & a_T & -a_L & 0 & \frac{\varepsilon_0}{2} & a_T & -a_L & \cdot \cr
\cdot & 0 & 0 & a_L & a_T & -\frac{\varepsilon_0}{2}+\omega & 0 & \cdot \cr
\cdot & 0 & 0 & a_T & -a_L & 0 & \frac{\varepsilon_0}{2}+\omega & \cdot \cr
\cdot & \cdot & \cdot & \cdot & \cdot & \cdot & \cdot & \cdot
\end{pmatrix}.
\end{footnotesize}
\label{eqn:HF-mat}
\end{align}
Here, $a_L=\frac{1}{4}A_L(\theta)$ and $a_T=\frac{1}{4}A_T(\theta)$.
We describe the eigenstate of $\H_F$ in the following form:
\begin{align}
\H_F \ket{\lambda_\gamma} = \lambda_\gamma \ket{\lambda_\gamma},
\end{align}
where $\lambda_\gamma$ is the $\gamma$th eigenvalue termed the quasienergy.
After taking the time average over a long period, the transition probability from $\alpha$ to $\beta$ states is expressed as
\cite{Shirley-1965}
\begin{align}
\bar{P}_{\alpha\rightarrow \beta}(\theta,\varepsilon_0) =
\sum_k \sum_\gamma \left| \braket{\beta k|\lambda_\gamma} \braket{\lambda_\gamma|\alpha 0} \right|^2.
\label{eqn:P-Shirley}
\end{align}

\subsection{Calculated results}

\subsubsection{Weak coupling limit}

In the weak coupling region, the transition probability $\bar{P}_{E_1\rightarrow E_2}^{(n)}$
mediated by $n$ phonons ($n$-phonon process) can be expressed analytically,
as shown in Appendix\ref{appendix:weak}.
In the weak coupling limit $A\rightarrow 0$, it vanishes for $n\ge 3$, whereas it stays finite for $n=1$ and $n=2$.
At the fixed energy of $\varepsilon_0=\omega$ and $\varepsilon_0=2\omega$,
the explicit forms of the $\theta$ dependence are given as
\begin{align}
&\bar{P}_{E_1\rightarrow E_2}^{(1)}(\theta,\varepsilon_0=\omega) = \frac{1}{2},
\label{eqn:e1} \\
&\bar{P}_{E_1\rightarrow E_2}^{(2)}(\theta,\varepsilon_0=2\omega)
= \frac{1}{2} \frac{\cos^2{2\theta}}{\cos^2{2\theta} + \left(\frac{2}{3}\right)^2 \sin^2{2\theta}}.
\label{eqn:e2}
\end{align}
Here, we used Eqs. (\ref{eqn:A}) and (\ref{eqn:P-ana-12}).

\begin{figure}[t]
\begin{center}
\includegraphics[width=7cm]{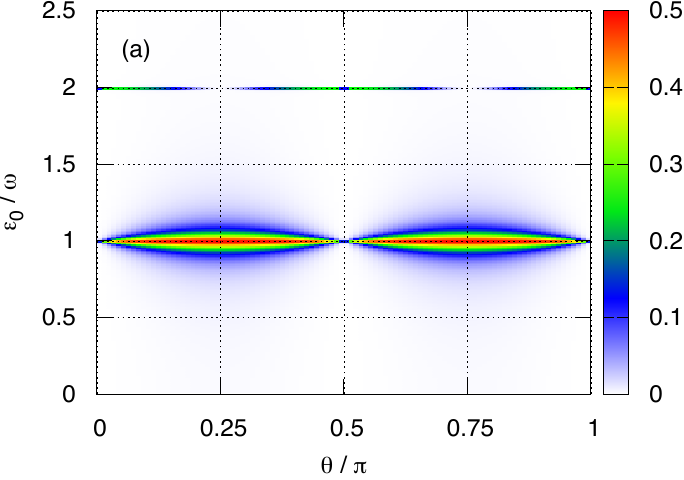}
\includegraphics[width=6.5cm]{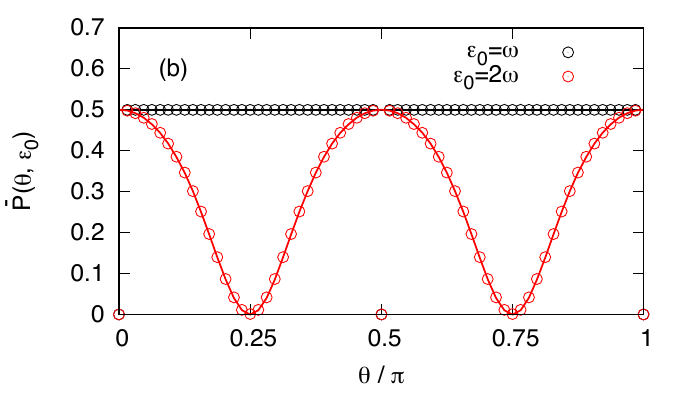}
\end{center}
\caption{
(Color online)
$\theta$ and $\varepsilon_0/\omega$ dependences of the transition probability
$\bar{P}_{E_1\rightarrow E_2}(\theta,\varepsilon_0)$.
The coupling constant in the weak coupling limit is chosen as $A=0.1$.
(a) Contour map of $\bar{P}_{E_1\rightarrow E_2}(\theta,\varepsilon_0)$.
(b) $\theta$ dependence at $\varepsilon_0=\omega$ and $\varepsilon_0=2\omega$.
The calculated transition probability vanishes at $\theta/\pi=0$, 0.5, and 1,
where $A_T(\theta)=0$ and no transition occurs.
The solid lines represent the analytic forms in Eqs. (\ref{eqn:e1}) and (\ref{eqn:e2}).
}
\label{fig:a=0.1}
\end{figure}

In Fig. \ref{fig:a=0.1}, we show the numerical result of the transition probability
given by Eq. (\ref{eqn:P-Shirley}) for $A=0.1$.
The contour map [Fig. \ref{fig:a=0.1}(a)] shows that the high transition probability
is concentrated in a narrow energy region at $\varepsilon_0 \simeq \omega$ and $\varepsilon_0 \simeq 2\omega$,
where they are resonance energies of the 1-phonon and 2-phonon processes, respectively.
The intensities stay finite even in the weak coupling limit, as expected.
The $\theta$ dependence of $\bar{P}_{E_1\rightarrow E_2}(\theta,\varepsilon_0)$ is shown in Fig. \ref{fig:a=0.1}(b)
at the resonance energies $\varepsilon_0=\omega$ and $\varepsilon_0=2\omega$.
The transition probability is constant at $\varepsilon_0=\omega$,
while it changes with $\theta$ at $\varepsilon_0=2\omega$,
except for $\theta/\pi = 0$, 0.5, and 1, at which it vanishes [$A_T(\theta)=0$].
In both cases, the numerical results of the $\theta$ dependences are well reproduced
by the analytic forms in Eqs. (\ref{eqn:e1}) and (\ref{eqn:e2}).

Next, we focus on the $\varepsilon_0=2\omega$ case.
Since the denominator in Eq. (\ref{eqn:e2}) does not significantly change with $\theta$,
the main $\theta$ dependence comes from the numerator.
The transition probability is proportional to $A_L^2(\theta)$ [see Eq. (\ref{eqn:analytic})]
and reflects the symmetry of the quadrupole for the longitudinal component.
Thus, by measuring the $\theta$ dependence of the resonance intensity at $\varepsilon_0=2\omega$,
we can obtain information on the quadrupole of the $S=1$ system.

\subsubsection{Weak coupling region}

\begin{figure}[t]
\begin{center}
\includegraphics[width=7cm]{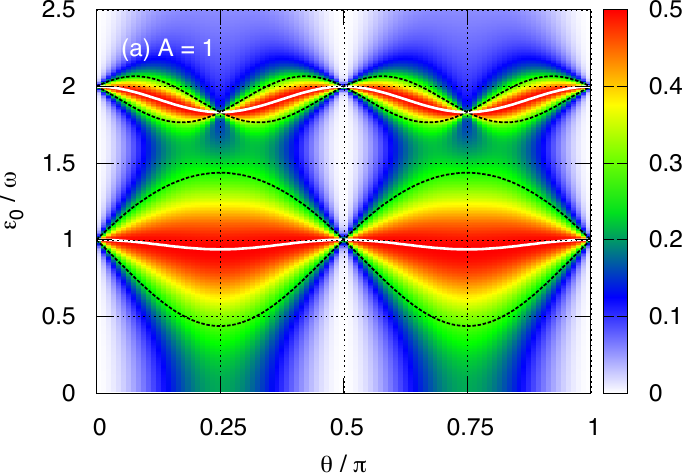}
\includegraphics[width=7cm]{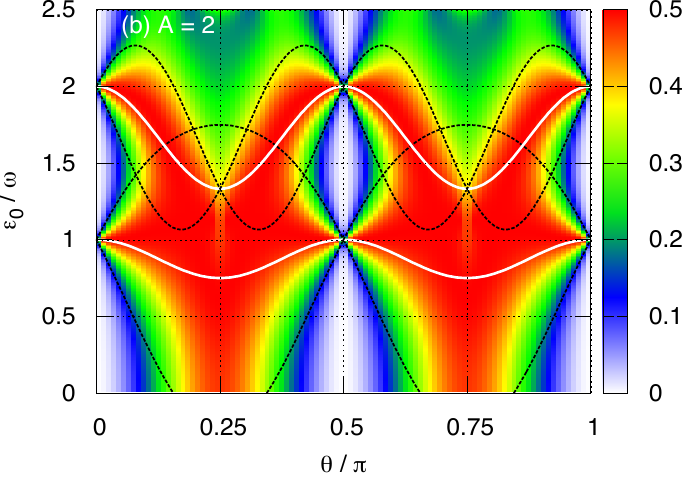}
\includegraphics[width=7cm]{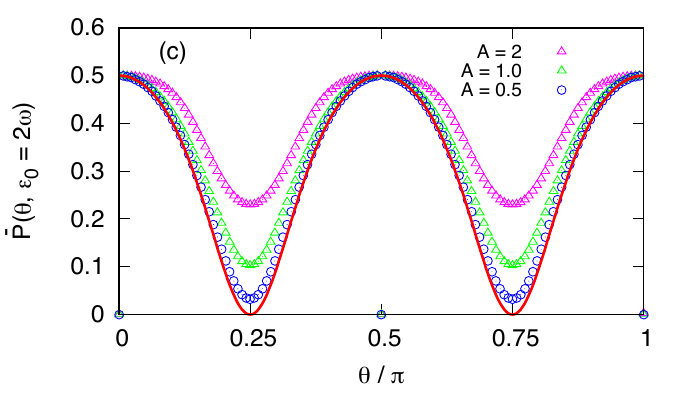}
\end{center}
\caption{
(Color online)
Transition probability for various coupling constants.
(a) Contour map of $\bar{P}_{E_1\rightarrow E_2}(\theta,\varepsilon_0)$ for $A=1$ in the weak coupling region.
The solid (white) lines represent the resonance energy
$\varepsilon_0 = n\omega + 2\delta_n$ for the $n$-phonon process ($n=1,2$).
The dotted (black) lines represent the broadening, i.e.,
$\varepsilon_0 = n\omega + 2\delta_n \pm 2|v_n|$.
(b) Contour map for $A=2$ in the strong coupling region.
(c) $\theta$ dependence at $\varepsilon_0=2\omega$ for $A=0.5$, $A=1$, and $A=2$.
The calculated transition probability vanishes since $A_T(\theta)=0$ at $\theta/\pi=0$, 0.5, and 1.
The solid (red) line is for the weak coupling limit given by Eq. (\ref{eqn:e2}).
}
\label{fig:strong}
\end{figure}

When the coupling $A$ increases, the transition probability deviates from the analytic formula for the weak coupling limit.
In Fig. \ref{fig:strong}(a), we show numerical results in the weak coupling region.
At $A=1$, the resonances become broad in energy for both the
1-phonon ($\varepsilon_0 \simeq \omega$) and 2-phonon ($\varepsilon_0 \simeq 2\omega$) processes [see Fig. \ref{fig:strong}(a)].
For the former process, the high intensity appears at around $\varepsilon_0=\omega$,
whereas it deviates from $\varepsilon_0=2\omega$ for the latter.

We analyze these results on the basis of the weak coupling theory given in Appendix\ref{appendix:weak}.
The transition probabilities for the $n$-phonon process are expressed as
[see Eqs. (\ref{eqn:P-ana}), (\ref{eqn:vn}), and (\ref{eqn:delta})]
\cite{Shirley-1965,Aravind-1984,Ho-1985,Son-2009,Koga-2020}
\begin{align}
\bar{P}_{E_1\rightarrow E_2}^{(n)}(\theta,\varepsilon_0) = \frac{1}{2}
\frac{(2v_{-n})^2}{(2v_{-n})^2 + (\varepsilon_0 - n\omega - 2\delta_n)^2},
\label{eqn:P-Lorentz}
\end{align}
where $2|v_{-n}|$ and $2\delta_n$ are shown in Table \ref{table:result}.
Equation (\ref{eqn:P-Lorentz}) indicates that the transition probability is expressed by a Lorentzian function.
The condition for the resonance in the weak coupling region is given by
\begin{align}
\varepsilon_0 = n\omega + 2\delta_n.
\label{eqn:resonance-s=1}
\end{align}
Here, $2\delta_n$ represents the level shift by the time-dependent periodic perturbation of the vibration.
\cite{Bloch-1940,note:Bloch-Siegert,Shirley-1965}
In Fig. \ref{fig:strong}(a), we show the resonance energy at $\varepsilon_0 = n\omega + 2\delta_n$
with the solid (white) line.
We can see that the shift is larger in
$\bar{P}_{E_1\rightarrow E_2}^{(2)}$ than in $\bar{P}_{E_1\rightarrow E_2}^{(1)}$ $(|\delta_2| > |\delta_1|)$.
For both $\bar{P}_{E_1\rightarrow E_2}^{(1)}$ and $\bar{P}_{E_1\rightarrow E_2}^{(2)}$,
the shifts are proportional to $A_T^2(\theta) \propto \sin^2{2\theta}$.
This means that the symmetry of the quadrupole coupled to the transverse component
can be detected by measuring the level shift.
The result in Fig. \ref{fig:strong}(a) shows that it is easier to measure the shift in $\bar{P}_{E_1\rightarrow E_2}^{(2)}$.

\begin{table}[t]
\caption{
Summary of level shift, broadening of the resonance, and transition probability at $\varepsilon_0=n\omega$ ($n=1,2$).
$A_L$ and $A_T$ are couplings for the longitudinal and transverse components defined in Eq. (\ref{eqn:A}), respectively,
where we omitted the $\theta$ dependence of $A_L(\theta)$ and $A_T(\theta)$ for abbreviation.
For the $n$-phonon process, the resonance condition is approximated
as the summation of the Van Vleck splitting and the level shift in Eq. (\ref{eqn:P-Lorentz}),
i.e., $\varepsilon_0 = \frac{1}{D}(g\mu_{\rm B}H)^2 = n\omega + 2\delta_n$,
where $2\delta_n$ is given for $n=1$ and $n=2$.
For $n\ge 2$, the level shift is given as $2\delta_n=-\frac{n}{4(n^2-1)}\frac{A_T^2}{\omega}$ [see Eq. (\ref{eqn:delta})].
Note that the transition probability vanishes for $A_T=0$.
This point is an exception among the listed formulae.
}
\begin{tabular}{ccc}
\hline
\hline
             & Level shift & $2\delta_1= -\frac{A_T^2}{16\omega}$ \\
$n=1$ & Broadening & $2|v_{-1}| = \frac{|A_T|}{2}$ \\
             & Probability & $\bar{P}_{E_1\rightarrow E_2}^{(1)}(\theta,\varepsilon_0=\omega) = \frac{1}{2} \frac{1}{1+\left(\frac{A_T}{8\omega}\right)^2}$ \\
\hline
             & Level shift & $2\delta_2 = -\frac{A_T^2}{6\omega}$ \\
$n=2$ & Broadening & $2|v_{-2}| = \frac{|A_T A_L|}{4\omega}$ \\
             & Probability & $\bar{P}_{E_1\rightarrow E_2}^{(2)}(\theta,\varepsilon_0=2\omega) = \frac{1}{2} \frac{A_L^2}{A_L^2+\left(\frac{2}{3}\right)^2 A_T^2}$ \\
\hline
\hline
\end{tabular}
\label{table:result}
\end{table}

In the Lorentzian function of Eq. (\ref{eqn:P-Lorentz}), $2|v_{-n}|$ causes the broadening.
In Fig. \ref{fig:strong}(a), we also show the lines $\varepsilon_0 = n\omega + 2\delta_n \pm 2|v_{-n}|$
for the broadening as dashed (black) lines,
on which $\bar{P}_{E_1\rightarrow E_2}^{(n)}(\theta,\varepsilon_0)=1/4$.
The broadening is proportional to $|A_T(\theta)| \propto |\sin{2\theta}|$ for $\bar{P}_{E_1\rightarrow E_2}^{(1)}$,
whereas it is proportional to $|A_T(\theta)A_L(\theta)| \propto |\sin{2\theta}\cos{2\theta}|\propto|\sin{4\theta}|$
for $\bar{P}_{E_1\rightarrow E_2}^{(2)}$.
These points are consistent with the numerical result.
Thus, the broadening of the resonance also carries information on the quadrupole degrees of freedom.
Note that the broadening factor $2|v_{-n}|$ also determines the intensity of the transition.
In the 2-phonon process, it is proportional to $|A_T A_L|$,
which indicates that both the longitudinal and transverse components are required for the transition.
\cite{Gromov-2000}

\subsubsection{Strong coupling region}

When the coupling is increased to $A=2$, the contour map of the transition probability changes markedly,
as shown in Fig. \ref{fig:strong}(b).
Since the 1-phonon and 2-phonon resonances become very broad and overlap, it is hard to distinguish them.
Nevertheless, the result with the weak coupling theory is useful
for providing a qualitative explanation for this strong coupling region.

We can see that the contour maps for $A=1$ and $A=2$ in Figs. \ref{fig:strong}(a) and \ref{fig:strong}(b), respectively,
are similar around $\theta/\pi=0$, 0.5, and 1, where $A_T(\theta)=A\sin{2\theta}$ almost vanishes
and the transition is forbidden at $A_T(\theta)=0$.
This shows that the weak coupling theory becomes valid for a small $A_T(\theta)$ even when $A$ is large,
as it was used for the photon-assisted magnetoacoustic resonance.
\cite{Koga-2020}

In the strong coupling region, the transition probability tends to take the maximum value
in wide energy $\varepsilon_0$ and field angle $\theta$ regions.
However, it suddenly decreases around the angles of $\theta/\pi=0$, $\theta/\pi=0.5$, and $\theta/\pi=1$,
as shown in Fig. \ref{fig:strong}(b).
This is a characteristic point in the strong coupling region,
and this feature can also be used to identify the quadrupole.
\cite{Koga-2020}

The transition probability on the $\varepsilon_0=\omega$ line is about 0.5,
and it does not change markedly with $\theta$ and $A$ [see Figs. \ref{fig:a=0.1}, \ref{fig:strong}(a), and \ref{fig:strong}(b)].
On the $\varepsilon_0=2\omega$ line, on the other hand, it strongly changes with $\theta$ as well as with $A$.
On the $\varepsilon_0=2\omega$ line, we show the $\theta$ dependence of $\bar{P}_{E_1\rightarrow E_2}(\theta,\varepsilon_0=2\omega)$
in Fig. \ref{fig:strong}(c) for $A=0.5$, $A=1$, and $A=2$.
By comparing these data, we can see that the $\theta$ dependence for a relatively large $A$
still keeps the characteristic feature of the weak coupling.

\section{$J=4$ System for $f^2$ Configuration in $O_h$ Symmetry}
\label{sec:oh}

\subsection{Effective Hamiltonian}

It is known that a non-Kramers doublet is realized in the $f^2$ configuration in $O_h$ point group symmetry.
It belongs to the $\Gamma_3$ ($E_g$) representation and the wave functions of the doublet are given by
\cite{Lea-1962}
\begin{align}
&\ket{+} = \frac{\sqrt{42}}{12} \left( \ket{4} + \ket{-4} \right) - \frac{\sqrt{15}}{6} \ket{0}, \cr
&\ket{-} = \frac{1}{\sqrt{2}} \left( \ket{2} + \ket{-2} \right).
\label{eqn:doublet}
\end{align}
Here, $\ket{m}$ represents the $J_z=m$ state for $J=4$.
Among the quadrupoles, only
$O_u=\frac{1}{\sqrt{3}}(3J_z^2-\bJ^2)=\frac{1}{\sqrt{3}}(2J_z^2-J_x^2-J_y^2)$ and $O_v=J_x^2-J_y^2$
have finite matrix elements within the doublet.
The matrix forms are given by
\begin{align}
O_u = \frac{8}{\sqrt{3}}
\begin{pmatrix}
1 & 0 \cr
0 & -1
\end{pmatrix},
~~~~~~
O_v = - \frac{8}{\sqrt{3}}
\begin{pmatrix}
0 & 1 \cr
1 & 0
\end{pmatrix}.
\label{eqn:ou-ov-oh}
\end{align}
As shown in Eq. (\ref{eqn:H-h-oh}), the effective Hamiltonian for the non-Kramers doublet
under a magnetic field is expressed in the following form:
\begin{align}
\H_{\rm eff}
= - c \left[ \frac{2h_z^2 - h_x^2 - h_y^2}{\sqrt{3}} O_u + (h_x^2 - h_y^2) O_v \right].
\label{eqn:H-eff-oh-0}
\end{align}
Here, $c$ is an arbitrary coupling constant, and we retained only the $O_u$ and $O_v$ quadrupoles.
Since $O_u$ and $O_v$ belong to the same $\Gamma_3$ representation in the $O_h$ symmetry,
the coupling constant $c$ is common to them.
As shown in Eq. (\ref{eqn:H-oh-2}), $O_u$ and $O_v$ couple to
$\frac{1}{\sqrt{3}}(2\varepsilon_{zz}-\varepsilon_{xx}-\varepsilon_{yy})$
and $\varepsilon_{xx}-\varepsilon_{yy}$ strains, respectively.

\subsection{$\bH\perp [111]$}

First, we study a case where the magnetic field is applied perpendicular to a threefold symmetrical direction ([111] direction).
For this, we introduce an $x'y'z'$ coordinate, where the $z'$-axis is taken
antiparallel to the [111] direction ($[\bar{1}\bar{1}\bar{1}]$ direction).
The $y'$-axis is chosen parallel to the $[\bar{1}10]$ direction.
Their unit vectors are given by $\be_{z'}=(-1,-1,-1)/\sqrt{3}$ and $\be_{y'}=(-1,1,0)/\sqrt{2}$, respectively.
The unit vector for the $x'$-axis is then defined by $\be_{x'}=\be_{y'}\times\be_{z'}=(-1,-1,2)/\sqrt{6}$.
On this basis, the magnetic field is expressed as
\begin{align}
\bh &= h \left( \cos\theta \be_{x'} + \sin\theta \be_{y'} \right) \cr
&= \sqrt{\frac{2}{3}} h
\left( \cos\left(\theta+\frac{2}{3}\pi\right), \cos\left(\theta-\frac{2}{3}\pi\right), \cos\theta \right).
\end{align}
Under this field,
$\frac{1}{\sqrt{3}}(2h_z^2-h_x^2-h_y^2)=\frac{h^2}{\sqrt{3}}\cos2\theta$
and $h_x^2-h_y^2=\frac{h^2}{\sqrt{3}}\sin2\theta$.
Then, the effective Hamiltonian in Eq. (\ref{eqn:H-eff-oh-0}) reduces to
\begin{align}
\H_{\rm eff}
&= -\frac{8}{3} c h^2
\begin{pmatrix}
\cos2\theta & -\sin2\theta \cr
-\sin2\theta & -\cos2\theta
\end{pmatrix}.
\label{eqn:H-eff-oh-111}
\end{align}

The effective Hamiltonian can be diagonalized as
\begin{align}
U^\dagger \H_{\rm eff} U
= - \frac{16}{3}ch^2 \frac{1}{2}
\begin{pmatrix}
1 & 0 \cr
0 & -1
\end{pmatrix},
\label{eqn:H-eff-oh-111-diag}
\end{align}
with
\begin{align}
&U =
\begin{pmatrix}
\cos\theta & \sin\theta \cr
-\sin\theta & \cos\theta
\end{pmatrix}.
\end{align}
Equation (\ref{eqn:H-eff-oh-111-diag}) indicates that the Van Vleck splitting is $(16/3)ch^2$.
It is isotropic (constant) with respect to $\theta$.
In the diagonalized basis, the quadrupole operators are transformed as
\begin{align}
&\tilde{O}_u = U^\dagger O_u U
= \frac{8}{\sqrt{3}}
\begin{pmatrix}
\cos{2\theta} & \sin{2\theta} \cr
\sin{2\theta} & -\cos{2\theta}
\end{pmatrix}, \cr
&\tilde{O}_v = U^\dagger O_v U
= \frac{8}{\sqrt{3}}
\begin{pmatrix}
\sin{2\theta} & -\cos{2\theta} \cr
-\cos{2\theta} & -\sin{2\theta}
\label{eqn:O-uv}
\end{pmatrix}.
\end{align}
We note that the transformed operators in Eq. (\ref{eqn:O-uv})
have the same form as those in Eq. (\ref{eqn:O-2}) in the $S=1$ case.
There is the following correspondence between the two cases:
$(O_u,O_v)_{O_h}\leftrightarrow (O_v,O_{xy})_{S=1}$.
The coefficient $8/\sqrt{3}$ in Eq. (\ref{eqn:O-uv}) can be absorbed
in the coupling constant between quadrupoles and strains.
For the diagonalized effective Hamiltonian in Eq. (\ref{eqn:H-eff-oh-111-diag}),
there is the following correspondence between the two cases:
$(16/3)c h^2 \leftrightarrow h^2/D (= \varepsilon_0)$.
Therefore, the model for the quadrupole resonance under the field $\bH\perp [111]$ in $O_h$ symmetry
can be mapped to that in the $S=1$ case under the field $\bH\perp z$.

\subsection{$\bH\perp z$}
\label{sec:j=4-oh}

When the magnetic field is applied in the $xy$-plane ($\bH\perp z$-axis),
the effective Hamiltonian in Eq. (\ref{eqn:H-eff-oh-0}) reduces to
\begin{align}
\H_{\rm eff}
&= - c \left[ \frac{- h_x^2 - h_y^2}{\sqrt{3}} O_u + (h_x^2 - h_y^2) O_v \right] \cr
&= \frac{8}{3} c h^2
\begin{pmatrix}
1 & \sqrt{3} \cos{2\theta} \cr
\sqrt{3} \cos{2\theta} & -1
\end{pmatrix}.
\label{eqn:H-eff-oh}
\end{align}
Here, we used $\bh=(h_x,h_y,0)=h(\cos\theta,\sin\theta,0)$.
The effective Hamiltonian can be diagonalized as
\begin{align}
U^\dagger \H_{\rm eff} U
= - c(\theta) h^2 \frac{1}{2}
\begin{pmatrix}
1 & 0 \cr
0 & -1
\end{pmatrix},
\end{align}
with
\begin{align}
&U =
\begin{pmatrix}
-\sin\phi & \cos\phi \cr
\cos\phi & \sin\phi
\end{pmatrix}, \cr
&\begin{pmatrix}
\cos{2\phi} \cr
\sin{2\phi}
\end{pmatrix}
= \frac{1}{\sqrt{ 1 + 3\cos^2{2\theta} }}
\begin{pmatrix}
1 \cr
\sqrt{3} \cos{2\theta}
\end{pmatrix}, \cr
&c(\theta) = \frac{16}{3} c \sqrt{ 1 + 3\cos^2{2\theta} } = \frac{16}{3} c \sqrt{ \frac{5}{2} + \frac{3}{2}\cos{4\theta} }.
\label{eqn:oh-c}
\end{align}
In the diagonalized basis, the quadrupole operators are transformed as
\begin{align}
&\tilde{O}_u = U^\dagger O_u U
= \frac{8}{\sqrt{3}}
\begin{pmatrix}
-\cos{2\phi} & -\sin{2\phi} \cr
-\sin{2\phi} & \cos{2\phi}
\end{pmatrix} \cr
&~~~
= \frac{8}{\sqrt{3}} \frac{1}{\sqrt{ 1 + 3\cos^2{2\theta} }}
\begin{pmatrix}
-1 & -\sqrt{3} \cos{2\theta} \cr
-\sqrt{3} \cos{2\theta} & 1
\end{pmatrix}, \cr
&\tilde{O}_v = U^\dagger O_v U
= \frac{8}{\sqrt{3}}
\begin{pmatrix}
\sin{2\phi} & -\cos{2\phi} \cr
-\cos{2\phi} & -\sin{2\phi}
\end{pmatrix} \cr
&~~~
= \frac{8}{\sqrt{3}} \frac{1}{\sqrt{ 1 + 3\cos^2{2\theta} }}
\begin{pmatrix}
\sqrt{3} \cos{2\theta} & -1 \cr
-1 & -\sqrt{3} \cos{2\theta}
\end{pmatrix}.
\end{align}

In the presence of the vibration coupled to the $O_u$ and $O_v$ quadrupoles,
the time-dependent effective Hamiltonian is given by
\begin{align}
\H_{\rm eff}(t)
&=
-  \frac{c(\theta)h^2}{2}
\begin{pmatrix}
1 & 0 \cr
0 & -1
\end{pmatrix}
+ \frac{1}{2}
\begin{pmatrix}
A_L(\theta) & A_T(\theta) \cr
A_T(\theta) & -A_L(\theta)
\end{pmatrix}
\cos{\omega t},
\end{align}
with
\begin{align}
\begin{pmatrix}
A_L(\theta) \cr
A_T(\theta)
\end{pmatrix}
&= \frac{8}{\sqrt{3}} \frac{1}{\sqrt{ 1 + 3\cos^2{2\theta} }} \cr
&\times
\left[ A_u
\begin{pmatrix}
-1 \cr
- \sqrt{3} \cos{2\theta}
\end{pmatrix}
+ A_v
\begin{pmatrix}
\sqrt{3} \cos{2\theta} \cr
-1
\end{pmatrix}
\right].
\label{eqn:ou-ov}
\end{align}
Here, ($A_u,A_v$) represent the coupling constants between ($O_u,O_v$) quadrupoles and 
$[\frac{1}{\sqrt{3}}(2\varepsilon_{zz}-\varepsilon_{xx}-\varepsilon_{yy}),\varepsilon_{xx}-\varepsilon_{yy}]$
strains, respectively.

\subsection{Numerical results}

\begin{figure}[t]
\begin{center}
\hspace{-0.95cm}
\includegraphics[width=6.5cm]{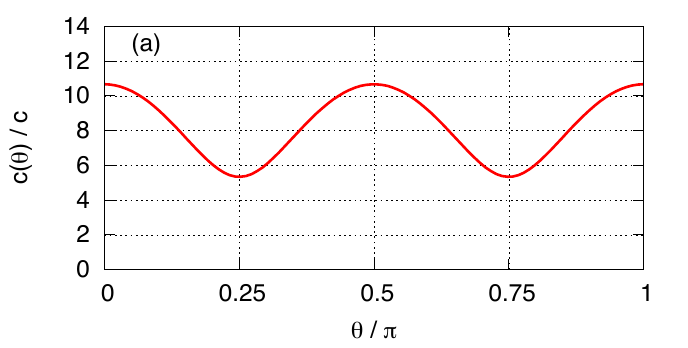}
\includegraphics[width=7.5cm]{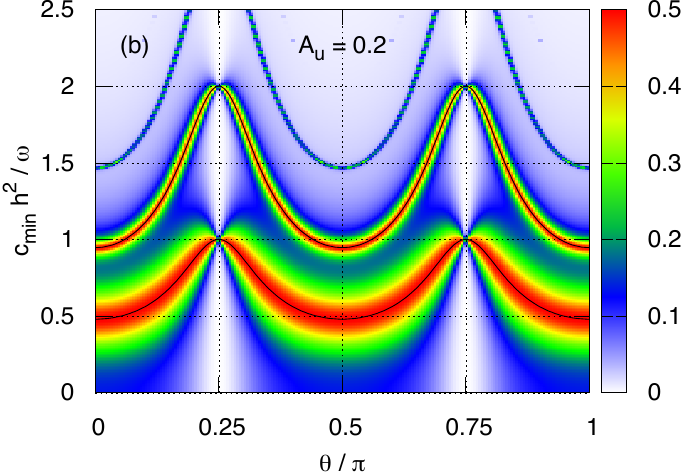}
\includegraphics[width=7.5cm]{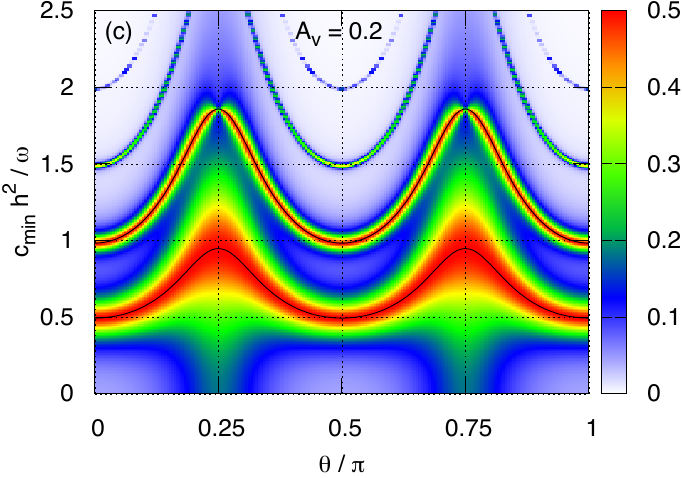}
\end{center}
\caption{
(Color online)
(a) $\theta$ dependence of $c(\theta)$ in $O_h$ symmetry [see the definition in Eq. (\ref{eqn:oh-c})].
The Van Vleck splitting is given by $c(\theta)h^2$ under the field $h=g\mu_{\rm B}H$ in the $xy$-plane.
(b) Contour map of the transition probability for $A_u=0.2$ ($A_v=0$).
The solid (black) lines are for the resonance magnetic field given by Eq. (\ref{eqn:resonance-oh}).
(c) Contour map of the transition probability for $A_v=0.2$ ($A_u=0$).
}
\label{fig:oh}
\end{figure}

We first show the $\theta$ dependence of $c(\theta)$ in Fig. \ref{fig:oh}(a).
The Van Vleck splitting is given by $c(\theta)h^2$.
The resonance magnetic field is anisotropic with the field direction.
Under a fixed field $h^2$, the splitting becomes large for a large $c(\theta)$.
In other words, to have a fixed splitting, a strong field $h^2$ is required for a small $c(\theta)$.

The transition probability is given as a function of both $\theta$ and $h^2$.
As in the $S=1$ case of Eq. (\ref{eqn:resonance-s=1}),
the resonance condition for the $n$-phonon process in the weak coupling limit is given by
\begin{align}
c(\theta)h^2 = n\omega + 2\delta_n,
\end{align}
where $c(\theta)h^2$ plays the role of $\varepsilon_0$ in Eq. (\ref{eqn:resonance-s=1}).
The $2\delta_n$ term represents the level shift, which is summarized in Table \ref{table:result}.
For the field dependence, we chose $c_{\rm min}h^2/\omega$ instead of $h^2$.
Here, $c_{\rm min}$ is the minimum value of $c(\theta)$ ($c_{\rm min}=16c/3$).
The resonance magnetic field, under which the resonance condition is satisfied, is then expressed as
\begin{align}
\frac{c_{\rm min}h^2}{\omega} = \left( n + \frac{2\delta_n}{\omega} \right) \frac{c_{\rm min}}{c(\theta)}.
\label{eqn:resonance-oh}
\end{align}

In Figs. \ref{fig:oh}(b) and \ref{fig:oh}(c), we show the numerical results of the transition probability
for the $O_u$ and $O_v$ quadrupoles, respectively.
We also show the resonance magnetic field of Eq. (\ref{eqn:resonance-oh}) with solid (black) lines for $n=1$ and $n=2$.
As expected, a high intensity appears near the line.
The lines are slightly different between Figs. \ref{fig:oh}(b) and \ref{fig:oh}(c).
This is caused by the level shift proportional to $-A_T^2(\theta)$ (see Table \ref{table:result}),
where the transverse component $A_T(\theta)$ is different
between the $O_u$ and $O_v$ cases [see Eq. (\ref{eqn:ou-ov})].
This difference also appears as the broadening effect,
which is proportional to $|A_T(\theta)|$ (see Table \ref{table:result}).
For $O_v$, $|A_T(\theta)|$ is large at $\theta=\pi/4$ and this leads to the large broadening and the level shift.
For $O_u$, in contrast, $|A_T(\theta)|=0$ and there is no broadening and no level shift at $\theta=\pi/4$.

Thus, by measuring the magnetoacoustic quadrupole resonance,
we can confirm the non-Kramers doublet which is closely associated with the $O_h$ symmetry.

\section{$J=4$ System for $f^2$ Configuration in $D_{4h}$ Symmetry}
\label{sec:d4h}

\subsection{Effective Hamiltonian}

For the $J=4$ system in a tetragonal $D_{4h}$ symmetry,
there are two doublet states belonging to the $\Gamma_5$ ($E_g$) representation.
We term them as $\Gamma_5^{(1)}$ and $\Gamma_5^{(2)}$, whose energy eigenvalues are different.
Since the energy of the $\Gamma_5$ doublet is not split linearly with the field applied in the $xy$-plane,
we can regard it as a non-Kramers doublet under the field $\bH\perp z$.
For the $\Gamma_5^{(1)}$ doublet, the wave functions are given by
\cite{Koga-1995,Kusunose-2011}
\begin{align}
\ket{+} = \alpha \ket{-3} + \beta \ket{1},~~~~~~
\ket{-} = \alpha \ket{3} + \beta \ket{-1}.
\label{eqn:doublet-d4h}
\end{align}
Here, $\alpha$ and $\beta$ are real constants satisfying $\alpha^2+\beta^2=1$.
We express them as
\begin{align}
\alpha = \sin\varphi,~~~~~~
\beta = \cos\varphi.
\label{eqn:alpha-beta}
\end{align}
Since the wave functions for the $\Gamma_5^{(2)}$ doublet are orthogonal to those for the $\Gamma_5^{(1)}$ doublet,
the wave functions for the $\Gamma_5^{(2)}$ doublet are given by Eq. (\ref{eqn:doublet-d4h})
with the replacement $\varphi\rightarrow \varphi+\pi/2$ in Eq. (\ref{eqn:alpha-beta}).
Thus, we can treat both doublet cases by changing $\varphi$ within $0 \le \varphi \le \pi$.

Among the quadrupoles, only $O_v=J_x^2-J_y^2$ and $O_{xy}=J_x J_y + J_y J_x$ have finite matrix elements
within the doublet in Eq. (\ref{eqn:doublet-d4h}).
The matrix forms are given by
\begin{align}
O_v = a
\begin{pmatrix}
0 & 1 \cr
1 & 0
\end{pmatrix},~~~~~~
O_{xy} = b
\begin{pmatrix}
0 & -i \cr
i & 0
\end{pmatrix},
\end{align}
with
\begin{align}
a = 10\beta^2 + 6\sqrt{7}\alpha\beta,~~~~~~
b = 10\beta^2 - 6\sqrt{7}\alpha\beta.
\label{eqn:ab}
\end{align}
The effective Hamiltonian is then expressed in the following form [see Eq. (\ref{eqn:H-h-d4h})]:
\begin{align}
\H_{\rm eff}
&= - c_v (h_x^2 - h_y^2) O_v - c_{xy} 2h_x h_y O_{xy} \cr
&= - c h^2
\begin{pmatrix}
0 & \tilde{a} - i \tilde{b} \cr
\tilde{a} + i \tilde{b} & 0
\end{pmatrix}.
\label{eqn:H-eff-d4h}
\end{align}
Here, $c_v$ and $c_{xy}$ are arbitrary constants.
Since the $O_v$ and $O_{xy}$ quadrupoles belong to different irreducible representations in the $D_{4h}$ symmetry,
$c_v$ and $c_{xy}$ can be different.
In Eq. (\ref{eqn:H-eff-d4h}), $c=\sqrt{c_v^2+c_{xy}^2}$ and we introduced
\begin{align}
\tilde{a} = \frac{c_v}{c} a,~~~~~~
\tilde{b} = \frac{c_{xy}}{c} b.
\label{eqn:ab-2}
\end{align}
Since the $O_v$ and $O_{xy}$ quadrupoles belong to the $\Gamma_3$ ($B_{1g}$) and $\Gamma_4$ ($B_{2g}$) representations,
they couple to the $\varepsilon_{xx}-\varepsilon_{yy}$ and $2\varepsilon_{xy}$ strains, respectively [see Eq. (\ref{eqn:H-d4h-2})].

The effective Hamiltonian can be diagonalized as
\begin{align}
U^\dagger \H_{\rm eff} U
= - \tilde{c}(\theta,\varphi) h^2 \frac{1}{2}
\begin{pmatrix}
1 & 0 \cr
0 & -1
\end{pmatrix},
\end{align}
with
\begin{align}
&U = \frac{1}{\sqrt{2}}
\begin{pmatrix}
i e^{-i\phi} & - e^{-i\phi} \cr
i e^{i\phi} & e^{i\phi}
\end{pmatrix}, \cr
&\begin{pmatrix}
\cos{2\phi} \cr
\sin{2\phi}
\end{pmatrix}
= \frac{1}{\sqrt{ \tilde{a}^2\cos^2{2\theta} + \tilde{b}^2\sin^2{2\theta} } }
\begin{pmatrix}
\tilde{a} \cos{2\theta} \cr
\tilde{b} \sin{2\theta}
\end{pmatrix}, \cr
&\tilde{c}(\theta,\varphi) = c \sqrt{ \frac{1}{2}( \tilde{a}^2 + \tilde{b}^2 ) + \frac{1}{2}( \tilde{a}^2 - \tilde{b}^2 ) \cos{4\theta} }.
\label{eqn:d4h-c}
\end{align}
In the diagonalized basis, the quadrupole operators are transformed as
\begin{align}
&\tilde{O}_v = U^\dagger O_v U
=
\begin{pmatrix}
\cos{2\phi} & \sin{2\phi} \cr
\sin{2\phi} & -\cos{2\phi}
\end{pmatrix} \cr
&\tilde{O}_{xy} = U^\dagger O_{xy} U
=
\begin{pmatrix}
\sin{2\phi} & -\cos{2\phi} \cr
-\cos{2\phi} & -\sin{2\phi}
\end{pmatrix}.
\end{align}

In the presence of the vibration coupled to the $O_v$ and $O_{xy}$ quadrupoles,
the time-dependent effective Hamiltonian is given by
\begin{align}
\H_{\rm eff}(t)
&=
-  \frac{\tilde{c}(\theta,\varphi)h^2}{2}
\begin{pmatrix}
1 & 0 \cr
0 & -1
\end{pmatrix}
+ \frac{1}{2}
\begin{pmatrix}
A_L(\theta) & A_T(\theta) \cr
A_T(\theta) & -A_L(\theta)
\end{pmatrix}
\cos{\omega t},
\end{align}
with
\begin{align}
\begin{pmatrix}
A_L(\theta) \cr
A_T(\theta)
\end{pmatrix}
= A_v
\begin{pmatrix}
\cos{2\phi} \cr
\sin{2\phi}
\end{pmatrix}
+ A_{xy}
\begin{pmatrix}
\sin{2\phi} \cr
-\cos{2\phi}
\end{pmatrix}.
\label{eqn:ov}
\end{align}
Here, ($A_v,A_{xy}$) represent the coupling constants between the ($O_v,O_{xy}$) quadrupoles
and ($\varepsilon_{xx}-\varepsilon_{yy},2\varepsilon_{xy}$) strains, respectively.

\subsection{Numerical results}

\begin{figure}[t]
\begin{center}
\hspace{-5.5mm}
\includegraphics[width=6.5cm]{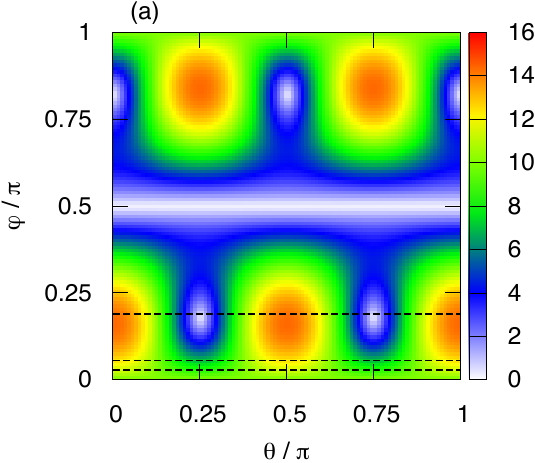}
\includegraphics[width=7.5cm]{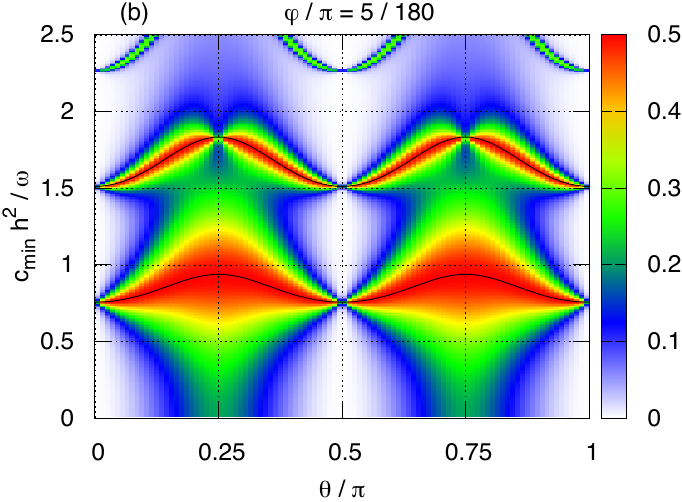}
\includegraphics[width=7.5cm]{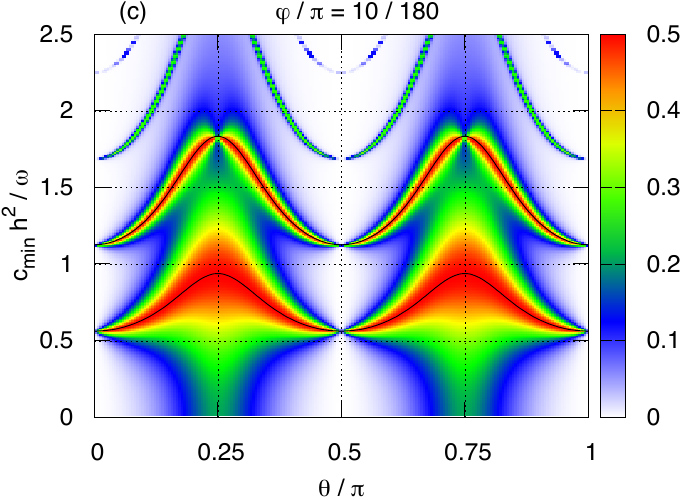}
\end{center}
\caption{
(Color online)
(a) $\theta$ and $\varphi$ dependences of $\tilde{c}(\theta,\varphi)/c$
in $D_{4h}$ symmetry [see the definition in Eq. (\ref{eqn:d4h-c})].
The dashed (black) lines from the bottom are for
$\varphi/\pi=5/180$, $\varphi/\pi=10/180$,
and $\varphi/\pi=\tan^{-1}[5/(3\sqrt{7})]/\pi\simeq 0.189~(\simeq 32.2~{\rm deg})$.
(b) Contour map of the transition probability for $\varphi/\pi=5/180$ with $A_v=1$ ($A_{xy}=0$).
For the $y$-axis, $c_{\rm min}=cb$ represents the minimum value of $\tilde{c}(\theta,\varphi)$ with respect to $\theta$.
The solid (black) lines are for the resonance magnetic field given by Eq. (\ref{eqn:resonance-oh})
with the replacement $c(\theta)\rightarrow \tilde{c}(\theta,\varphi)$.
(c) Contour map of the transition probability for $\varphi/\pi=10/180$ with $A_v=1$ ($A_{xy}=0$).
}
\label{fig:d4h}
\end{figure}

In Eq. (\ref{eqn:d4h-c}), $\cos{2\phi}$ and $\sin{2\phi}$ depend on both $\tilde{a}$ and $\tilde{b}$.
In turn, $\tilde{a}$ and $\tilde{b}$ depend on $c_v$, $c_{xy}$, $\alpha$, and $\beta$.
For simplicity, we assume that $c_v=c_{xy}=c$.
In this case, $\tilde{a}=a$ and $\tilde{b}=b$ [see Eq. (\ref{eqn:ab-2})],
which depend on $\alpha$ and $\beta$ [see Eq. (\ref{eqn:ab})].
When $\varphi=0$ in Eq. (\ref{eqn:alpha-beta}), $(\alpha,\beta)=(0,1)$, $\tilde{a}=\tilde{b}=10$,
and $(\cos{2\phi},\sin{2\phi})=(\cos{2\theta},\sin{2\theta})$.
The Van Vleck splitting $\tilde{c}(\theta,\varphi)h^2$ becomes isotropic [$\tilde{c}(\theta,\varphi)h^2=10ch^2$],
as in the $S=1$ case.
Thus, $\tilde{c}$ generally depends on both $\theta$ and $\varphi$.
We show the $\theta$ and $\varphi$ dependences of $\tilde{c}(\theta,\varphi)$ in Fig. \ref{fig:d4h}(a).
For $\varphi=0$ (or $\pi$), $\tilde{c}(\theta,\varphi)$ is independent of $\theta$.
For $\varphi=\pi/2$, we have $\tilde{c}(\theta,\varphi)=0$, since $\tilde{a}=\tilde{b}=0$ ($\beta=0$).
Except for these cases, $\tilde{c}(\theta,\varphi)$ depends on $\theta$.
The $\theta$ dependence becomes the most prominent when $\tilde{b}=0$ with $\tilde{a}\neq 0$.
This is realized for $\varphi= \tan^{-1}[5/(3\sqrt{7})]\simeq 0.189\pi~(\simeq 32.2~{\rm deg})$,
where $\tilde{c}(\theta,\varphi)=0$ at $\theta=\pi/4$.

We first consider a vibration coupled to the $O_v$ quadrupole, i.e., $A_{xy}=0$ in Eq. (\ref{eqn:ov}).
In Fig. \ref{fig:d4h}(b), we show the contour map of the transition probability for $\varphi/\pi=5/180$.
As in the $O_h$ case in Sect. \ref{sec:oh}, the $y$-axis is chosen as $c_{\rm min} h^2/\omega$,
where $c_{\rm min}=cb$ represents the minimum value of $\tilde{c}(\theta,\varphi)$ with respect to $\theta$.
For $\varphi=0$, there is no anisotropy in the Van Vleck splitting, as mentioned above.
For $\varphi/\pi=5/180$ (5 deg), the splitting becomes anisotropic in $\theta$.
Correspondingly, the resonance magnetic field shown with the solid (black) line in Fig. \ref{fig:d4h}(b) also becomes anisotropic.
This $\varphi$ corresponds to the first dashed line from the bottom in Fig. \ref{fig:d4h}(a).
When $\varphi$ is increased to $\varphi/\pi=10/180$ (10 deg),
the anisotropy becomes strong, as shown in Fig. \ref{fig:d4h}(c).
This corresponds to the second dashed line from the bottom in Fig. \ref{fig:d4h}(a).
In the limit of $\varphi\rightarrow \tan^{-1}[5/(3\sqrt{7})]~(\simeq 32.2~{\rm deg})$,
the anisotropy becomes maximum and an infinite value of the resonance field is required at $\theta=\pi/4$.
This corresponds to the third dashed line from the bottom in Fig. \ref{fig:d4h}(a),
on which $\tilde{c}(\varphi,\theta)=0$ at $\theta=\pi/4$ and there is no Van Vleck splitting there.
For $\varphi/\pi >1/2$, $\tilde{c}(\theta,\varphi)$ takes a minimum value at $\theta=0$, $0.5$, and 1,
and the resonance field becomes maximum there.
By measuring these features on the $\theta-H^2$ plane and by evaluating the value of $\varphi$,
we can determine the wave functions of the non-Kramers doublet.

Next, we discuss a vibration coupled to the $O_{xy}$ quadrupole.
Under the replacement of $A_v\rightarrow A_{xy}$ in Eq. (\ref{eqn:ov}), we can see the following relation:
$(\cos{2\phi},\sin{2\phi})\rightarrow(\sin{2\phi},-\cos{2\phi})$.
This can be realized by the replacement of $\theta\rightarrow \theta-\pi/4$
and $(\tilde{a},\tilde{b})\rightarrow (\tilde{b},\tilde{a})$ in Eq. (\ref{eqn:d4h-c}).
The second replacement is realized by $\varphi\rightarrow \pi-\varphi$ [see Eqs. (\ref{eqn:alpha-beta}) and (\ref{eqn:ab})].
Therefore, we can obtain the result for the $O_{xy}$ quadrupole
from that for $O_v$ by replacing $(\theta,\varphi)\rightarrow (\theta-\pi/4,\pi-\varphi)$.

\section{Probing Quadrupole Order}
\label{sec:order}

In the previous sections, we have seen that the quadrupole degrees of freedom
are detectable by means of the magnetoacoustic quadrupole resonance.
The same idea is applicable to observe quadrupole orders.
We demonstrate this for the non-Kramers doublet system in the $O_h$ symmetry with the following model:
\begin{align}
\H &= - c \left[ \frac{2h_z^2-h_x^2-h_y^2}{\sqrt{3}} O_u + (h_x^2-h_y^2) O_v \right] \cr
&~~~- J_{\rm eff}^u \braket{O_u} O_u - J_{\rm eff}^v \braket{O_v} O_v \cr
&~~~+ \frac{1}{2} ( A_u O_u + A_v O_v ) \cos\omega t.
\label{eqn:H-order}
\end{align}
Here, $J_{\rm eff}^\alpha(>0)$ represents an effective ferro-quadrupole intersite coupling for $O_\alpha$ ($\alpha=u,v$).
In the $O_h$ symmetry, $J_{\rm eff}^u=J_{\rm eff}^v$ is expected.
The matrix forms of the quadrupole operators are given by Eq. (\ref{eqn:ou-ov-oh}).
$\braket{O_\alpha}$ is the expectation value of the quadrupole operator at temperature $T$.
Since a ferro-quadrupole order is stabilized, the expectation value is common on all sites.
($A_u,A_v$) represent the coupling constants between ($O_u,O_v$) quadrupoles and
[$\frac{1}{\sqrt{3}}(2\varepsilon_{zz}-\varepsilon_{xx}-\varepsilon_{yy}),\varepsilon_{xx}-\varepsilon_{yy}$] strains, respectively.

\begin{figure*}[t]
\begin{center}
\includegraphics[width=6.6cm]{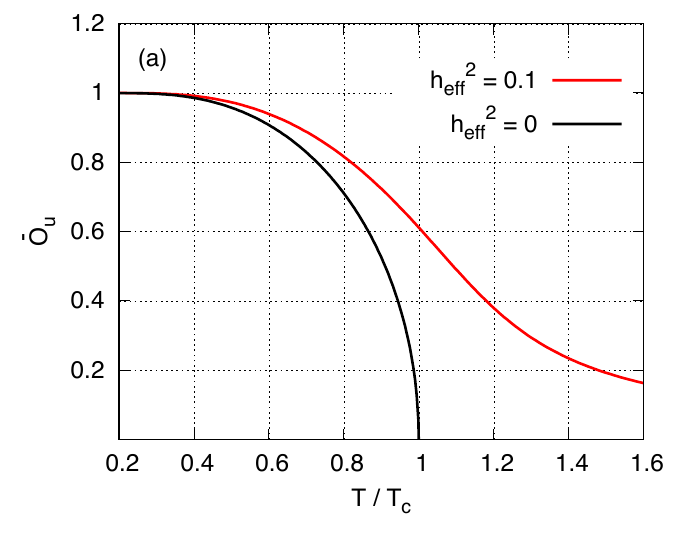}
\hspace{1.8cm}
\includegraphics[width=7.5cm]{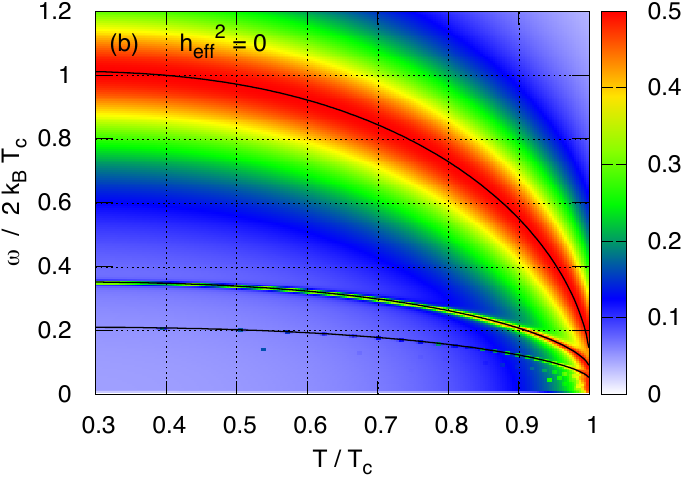}
\includegraphics[width=7.5cm]{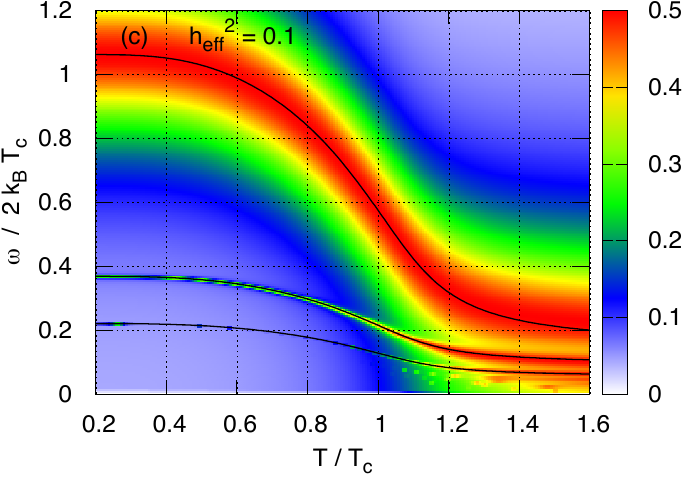}
\hspace{1cm}
\includegraphics[width=7.5cm]{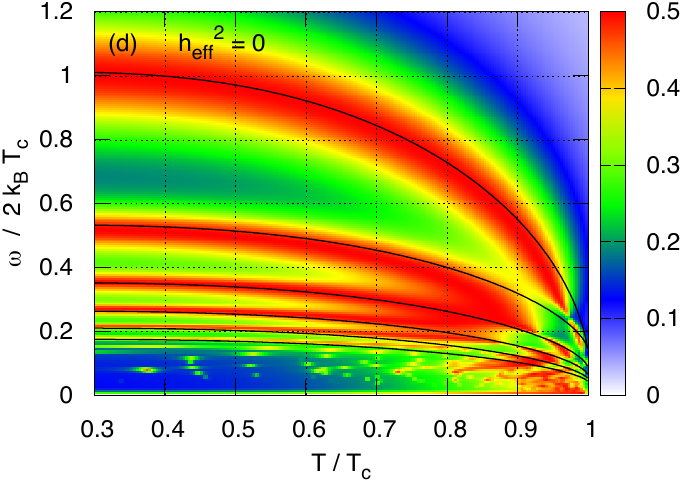}
\end{center}
\caption{
(Color online)
(a) Temperature dependence of the order parameter $\bar O_u=\braket{O_u}/(8/\sqrt{3})$
under the fields along the $z$-axis.
(b) Contour map of the transition probability on the $T-\omega$ plane for
$h_{\rm eff}^2=0~(h_{\rm eff}^2\rightarrow 0)$ and $(a_u,a_v)=(0,0.2)$.
The solid lines represent the resonance frequencies of the $n$-phonon process
given by Eq. (\ref{eqn:omega-n}) on the basis of the weak coupling theory.
The lines from the top are for $n=1$, $n=3$,  and $n=5$.
(c) For $h_{\rm eff}^2=0.1$ and $(a_u,a_v)=(0,0.2)$.
(d) For $h_{\rm eff}^2=0$ and
$(a_u,a_v)=(0.2\sqrt{3},\pm 0.2)$.
The transition probability does not depend on the signs of $a_u$ and $a_v$.
The lines from the top are for $n=1$, $n=2$, $n=3$, $n=4$, $n=5$, and $n=6$.
}
\label{fig:ou-order}
\end{figure*}

We first study a case where the $O_u$ quadrupole is stabilized,
assuming that an external magnetic field is applied in the $z$ direction.
For $\bh=(0,0,h_z)$, the magnetic field only couples to $O_u$, whereas $O_v$ is not induced by the field.
\cite{note:c}
On the basis of a conventional mean-field theory, the critical temperature is given as $k_{\rm B} T_c=(64/3)J_{\rm eff}^u$.
Here, the factor $64/3=(8/\sqrt{3})^2$ originates from the coefficient of the matrix form of $O_u$ in Eq. (\ref{eqn:ou-ov-oh}).
We then define the following dimensionless Hamiltonian for $\braket{O_v}=0$:
\begin{align}
\frac{\H}{k_{\rm B}T_c} = - \left( h_{\rm eff}^2 + 2\bar{O}_u \right) \frac{1}{2} \sigma_z
+ \frac{1}{2} ( a_u \sigma_z + a_v \sigma_x ) \cos\omega t.
\label{eqn:omega-solution}
\end{align}
Here, the Pauli matrices $\sigma_z$ and $\sigma_x$ are related to the quadrupole operators as
$\sigma_z\propto O_u$ and $\sigma_x\propto O_v$, respectively [see Eq. (\ref{eqn:ou-ov-oh})].
In Eq. (\ref{eqn:omega-solution}), $h_{\rm eff}^2\propto h_z^2$
and it represents the effective field coupled to $O_u$.
$\bar{O}_u=\braket{O_u}/(8/\sqrt{3})=\braket{\sigma_z}$ is a renormalized order parameter.
($a_u,a_v)=(1/k_{\rm B}T_c)(8/\sqrt{3})(A_u,A_v)$ represent the effective coupling constants
between the $(O_u,O_v)$ quadrupoles and the corresponding strains, respectively.

On the basis of the weak coupling theory, the resonance magnetic field for the $n$-phonon process
is given by [see Eq. (\ref{eqn:resonance-s=1}), Table \ref{table:result}, and Eq. (\ref{eqn:delta})]
\begin{align}
h_{\rm eff}^2 + 2\bar{O}_u = n \tilde\omega + \frac{2\tilde{\delta}_n}{\tilde\omega},
\label{eqn:omega}
\end{align}
where
\begin{align}
\tilde\omega = \frac{\omega}{k_{\rm B}T_c},~~~
2\tilde\delta_1= -\frac{a_v^2}{16},~~~
2\tilde\delta_n = -\frac{n a_v^2}{4(n^2-1)}.~~~(n\ge 2)
\end{align}
The $h_{\rm eff}^2 + 2\bar{O}_u$ term represents the energy splitting,
while $2\tilde{\delta}_n/\tilde\omega$ is the level shift.
In Eq. (\ref{eqn:omega}), the solution of the dimensionless $\tilde\omega$ for the $n$-phonon process is given by
\begin{align}
\tilde\omega = \frac{1}{2n}
\left[
h_{\rm eff}^2 + 2\bar{O}_u + \sqrt{ \left( h_{\rm eff}^2 + 2\bar{O}_u \right)^2-4n \left(2\tilde\delta_n\right)}
\right].
\label{eqn:omega-n}
\end{align}
In Fig. \ref{fig:ou-order}, the resonance frequencies for the $n$-phonon processes are shown with solid (black) lines.

We show the temperature dependence of the order parameter $\bar{O}_u$ in Fig. \ref{fig:ou-order}(a).
For $h_{\rm eff}^2=0~(h_{\rm eff}^2\rightarrow 0)$,
the order parameter spontaneously appears below $T_c$ and saturates at low temperatures.
For $h_{\rm eff}^2=0.1$, $\bar{O}_u$ is already induced above $T_c$  by the $h_z$ field
and abruptly develops near $T_c$.
The contour map of the transition probability on the $T-\omega$ plane is shown
in Fig. \ref{fig:ou-order}(b) for $h_{\rm eff}^2=0$ and $(a_u,a_v)=(0,0.2)$.
This combination of the coupling constants can be realized
by the vibration of the $\varepsilon_{xx}-\varepsilon_{yy}=\varepsilon_v$ strain
[see Eqs. (\ref{eqn:oh-u})--(\ref{eqn:H-oh-2})].
The energy splitting of the non-Kramers doublet is $2k_{\rm B}T_c\bar{O}_u$.
We can see that a high intensity appears for the 1-phonon process below $T_c$.
The resonance frequency increases with the development of the order parameter.
Observing the emergence of the spontaneous excitation gap below $T_c$ can be strong evidence of the quadrupole order.
Under a finite field along the $z$-axis, there is already a finite excitation gap (Van Vleck splitting) for $T>T_c$
owing to the induced $O_u$ moment, as shown in Fig. \ref{fig:ou-order}(c) for $h_{\rm eff}^2=0.1$.
The excitation gap, which abruptly increases near $T_c$, develops with decreasing temperature.

Note that the transition between the two states occurs
only for a finite transverse component ($a_v\neq 0$).
In the absence of the longitudinal component ($a_u=0$),
only the $n={\rm odd}$ phonon processes are possible, as studied by Shirley.
\cite{Shirley-1965}
This is due to the fact that the matrix elements for the $n={\rm even}$ process
are decoupled from those for the $n={\rm odd}$ process.
In the presence of both the transverse and longitudinal components,
all the $n$-phonon processes are coupled and the 2-phonon process can be seen,
\cite{Gromov-2000}
as demonstrated in Fig. \ref{fig:ou-order}(d) for
$(a_u,a_v)=(0.2\sqrt{3},\pm 0.2)$.
These combinations of the coupling constants can be realized by the vibration of the $\varepsilon_{zz}-\varepsilon_{yy}$
and $\varepsilon_{zz}-\varepsilon_{xx}$ strains, respectively
[see Eqs. (\ref{eqn:oh-u})--(\ref{eqn:H-oh-2})].
In both the $n={\rm odd}$ and $n={\rm even}$ cases,
the resonance frequency plotted with the solid lines in Fig. \ref{fig:ou-order}(d) well reproduces the numerical results.
This indicates that the formula for the level shift in Eq. (\ref{eqn:delta}) works for arbitrary $n$ values.

We can discuss a case of the $O_v$ order in parallel with the $O_u$ order studied above.
In this case, we assume a finite field $\bh\propto (\sqrt{2},0,1)$ to stabilize the $O_v$ order.
\cite{note:c2}
Then, the roles of $O_u$ and $O_v$ are interchanged,
and $a_u$ and $a_v$ work as the couplings to the transverse and longitudinal components, respectively.

In the absence of the external magnetic field, the $O_u$ and $O_v$ order parameters are degenerate.
In this case, the order parameter is a linear combination of $O_u$ and $O_v$.
This is also probed by using suitable combinations of the $\varepsilon_u$ and $\varepsilon_v$ strains.
Thus, the magnetoacoustic quadrupole resonance can be used to identify the symmetry of the quadrupole order parameter.

\section{Summary and Discussion}

In this paper, we first investigated details of the magnetoacoustic quadrupole resonance of an $S=1$ system
on the basis of the Floquet theory.
The $S=1$ system is a fundamental model of the two-level system coupled to the oscillating field
not only with the transverse (off-diagonal) component but also with the longitudinal (diagonal) one.
We derived the analytic form of the transition probability within the weak coupling theory,
which includes not only the 1-phonon process but also the multiphonon process.
The formula consists of the resonance energy with the level shift and the broadening factor.
It is applicable in the weak coupling region and also provides qualitative information even in the strong coupling region.

The theory for the $S=1$ model can be applied to realistic non-Kramers systems.
We focused on the $J=4$ system of the $f^2$ configuration in $O_h$ and $D_{4h}$ symmetries.
Since the acoustic wave couples to the non-Kramers doublet, which is a possible crystal-field ground state,
through the quadrupole degrees of freedom,
the magnetoacoustic resonance is useful as a microscopic probe of the quadrupole as follows.
The degeneracy of the doublet is lifted by an external magnetic field.
The excitation gap opens and changes with the rotation of the field direction,
which is characteristic of the anisotropic $O_h$ and $D_{4h}$ symmetries.
We can confirm the non-Kramers doublet by measuring dependences of the transition probability
on the direction and amplitude of the field.

When the non-Kramers system shows quadrupole ordering,
the degeneracy of the doublet is lifted even in the absence of an external magnetic field
owing to the spontaneous quadrupole field originating from the neighboring sites.
Since the splitting of the doublet increases with the development of the quadrupole order parameter,
the resonance frequency increases with the lowering of the temperature.
The resonance condition depends on the symmetries
of both the order parameter and the strain driven by the acoustic wave.
By analysis using the present theory, the quadrupole order can be confirmed by the magnetoacoustic resonance.

As an example of a specific material, the heavy-fermion superconductor URu$_2$Si$_2$
has been well known for a long time to have a puzzling issue of hidden order.
\cite{Mydosh-2011,Mydosh-2020}
A huge number of crystal-field models of $D_{4h}$ symmetry have been suggested for a U$^{4+}$ 5f$^2$ ion.
Among them, the $\Gamma_5$ non-Kramers doublet, which was discussed in Sect. 4,
is one possible candidate of the U ground state,
since it has both magnetic dipole and nonmagnetic quadrupole characters.
This was first indicated by a pioneering experimental study on dilute U alloys U$_x$Th$_{1-x}$Ru$_2$Si$_2$,
which show an abnormal metallic behavior associated with the multichannel Kondo effect.
\cite{Amitsuka-1994,Cox-1998}
However, the quadrupole ordering scenario for the hidden order
has been in disagreement with the resonant X-ray diffraction measurements thus far.
\cite{Amitsuka-2010, Walker-2011}
On the other hand, the large magnetic moment in the antiferromagnetic phase induced under a high pressure
may be due to an Ising-like magnetic moment of an U ion.
\cite{Amitsuka-2007,Wilson-2016}
It is strongly urged to reinvestigate the hidden quadrupole of the $\Gamma_5$-doublet origin
by using magnetoacoustic quadrupole resonance as a complementary probe to resonant X-ray diffraction.
The present analysis of the non-Kramers doublet for the $D_{4h}$ symmetry studied in Sect. \ref{sec:d4h}
may provide useful information on this.

Focusing on the non-Kramers doublet, we demonstrated how to identify the relevant quadrupole
by means of magnetoacoustic resonance.
We emphasize here that the present theory is applicable
not only to the non-Kramers doublet but also to other types of multiplet formed in a magnetic ion.
\cite{Koga-2020}
We hope that this work will draw attention to the use of magnetoacoustic resonance
for probing quadrupole degrees of freedom in quantum magnets.

\vspace{1mm}
{\footnotesize\paragraph{\footnotesize Acknowledgment}
This work was supported by JSPS KAKENHI Grant Number 17K05516.
}

\appendix
\setcounter{equation}{0}

\section{Matrices of $S=1$ Spin and Quadrupole Operators}
\label{appendix:matrix}

\subsection{$S=1$ spin operators}

The matrix forms of the $S=1$ spin operators are expressed as
\begin{align}
&S_x =
\begin{pmatrix}
0 & \frac{1}{\sqrt{2}} & 0 \cr
\frac{1}{\sqrt{2}} & 0 & \frac{1}{\sqrt{2}} \cr
0 & \frac{1}{\sqrt{2}} & 0
\end{pmatrix},~~~
S_y =
\begin{pmatrix}
0 & - \frac{i}{\sqrt{2}} & 0 \cr
\frac{i}{\sqrt{2}} & 0 & - \frac{i}{\sqrt{2}} \cr
0 & \frac{i}{\sqrt{2}} & 0
\end{pmatrix}, \cr
&S_z =
\begin{pmatrix}
1 & 0 & 0 \cr
0 & 0 & 0 \cr
0 & 0 & -1
\end{pmatrix}.
\label{eqn:S-mat}
\end{align}

\subsection{$S=1$ quadrupole operators}

The quadrupole operators for the $S=1$ spin are expressed as
\begin{align}
&O_{yz} = S_y S_z + S_z S_y
=
\begin{pmatrix}
0 & - \frac{i}{\sqrt{2}} & 0 \cr
\frac{i}{\sqrt{2}} & 0 & \frac{i}{\sqrt{2}} \cr
0 & -\frac{i}{\sqrt{2}} & 0
\end{pmatrix}, \cr
&O_{zx} = S_z S_x + S_x S_z
=
\begin{pmatrix}
0 & \frac{1}{\sqrt{2}} & 0 \cr
\frac{1}{\sqrt{2}} & 0 & -\frac{1}{\sqrt{2}} \cr
0 & -\frac{1}{\sqrt{2}} & 0
\end{pmatrix}, \cr
&O_{xy} = S_x S_y + S_y S_x
=
\begin{pmatrix}
0 & 0 & -i \cr
0 & 0 & 0 \cr
i & 0 & 0
\end{pmatrix}, \cr
&O_u = \frac{1}{\sqrt{3}} \left(3S_z^2 - \bS^2 \right)
=
\begin{pmatrix}
\frac{1}{\sqrt{3}} & 0 & 0 \cr
0 & -\frac{2}{\sqrt{3}} & 0 \cr
0 & 0 & \frac{1}{\sqrt{3}}
\end{pmatrix}, \cr
&O_v = S_x^2 - S_y^2
=
\begin{pmatrix}
0 & 0 & 1 \cr
0 & 0 & 0 \cr
1 & 0 & 0 \cr
\end{pmatrix}.
\label{eqn:O-mat}
\end{align}

\section{Fourth-Rank Matter Tensor}
\label{appendix:matter-tensor}

\subsection{Quadrupole--strain coupling}
\label{appendix:spin-lattice}

Interactions between spin (or angular momentum) and lattice strain are expressed by the following general form:
\cite{Watkins-1962,Donoho-1964}
\begin{align}
\H_\varepsilon = K_{ijkl} \varepsilon_{ij} J_k J_l.
\label{eqn:H-strain-spin}
\end{align}
Here,
\begin{align}
\varepsilon_{ij} = \frac{1}{2} \left( \frac{\partial u_i}{\partial x_j} + \frac{\partial u_j}{\partial x_i} \right)
\label{eqn:strain}
\end{align}
denotes the strain tensor with $\bm{u}=(u_x,u_y,u_z)$ as the displacement vector.
$J_k$ represents the $k$th component of the angular momentum operator.
The product of the angular momentum operators $J_k J_l$ is termed as the quadrupole.
$K_{ijkl}$ is the coefficient for the quadrupole--strain coupling.
Since the strain tensor is symmetric ($\varepsilon_{ij}=\varepsilon_{ji}$), the coefficient satisfies $K_{ijkl}=K_{jikl}$.
To keep the time-reversal symmetry, $K_{ijkl}$ must be a real value ($K_{ijkl}=K_{ijkl}^*$).
To satisfy the Hermitian nature of the Hamiltonian, $K_{ijkl}$ is symmetric with respect to $kl$ ($K_{ijkl}=K_{ijlk}$).
Since the Hamiltonian in Eq. (\ref{eqn:H-strain-spin}) must belong to a $\Gamma_1$ representation,
the coefficient $K_{ijkl}$ is determined so as to satisfy the invariance of the Hamiltonian under symmetry transformations.
Such $K_{ijkl}$ is known as a fourth-rank matter tensor and is classified on the basis of the point group symmetries.
\cite{Nye-1985,Nowick-1996,Powell-2010}
In the following subsections, we present the general forms of the quadrupole--strain coupling
for the $O_h$ and $D_{4h}$ point groups, which are studied in this paper.
In other point group symmetries,
note that the general forms of the quadrupole--strain coupling are also derived from the fourth-rank matter tensor.
We remark that the quadrupole--strain coupling can also be constructed
by the linear combination of the basis functions of irreducible representations.
\cite{Udvarhelyi-2018}

\subsubsection{$O_h$ symmetry}

For the $O_h$ point group, the quadrupole--strain coupling is expressed
on the basis of the fourth-rank matter tensor in the following general form:
\cite{Nye-1985,Nowick-1996,Powell-2010}
\begin{align}
\H_\varepsilon
&= K_{11} \left( \varepsilon_{xx} O_{xx} + \varepsilon_{yy} O_{yy} + \varepsilon_{zz} O_{zz} \right) \cr
&+ K_{12} \left[ \left( \varepsilon_{yy} + \varepsilon_{zz} \right) O_{xx}
                + \left( \varepsilon_{xx} + \varepsilon_{zz} \right) O_{yy}
                + \left( \varepsilon_{xx} + \varepsilon_{yy} \right) O_{zz} \right] \cr
&+ \frac{1}{2} K_{44} \left( 2\varepsilon_{yz} O_{yz} + 2\varepsilon_{zx} O_{zx} + 2\varepsilon_{xy} O_{xy} \right).
\label{eqn:H-oh}
\end{align}
Here, $K_{11}$, $K_{12}$, and $K_{44}$ are arbitrary real constants.
In Eq. (\ref{eqn:H-oh}), $O_{\alpha\alpha}=J_\alpha^2$
and $O_{\alpha\beta}=J_\alpha J_\beta + J_\beta J_\alpha$ ($\alpha\neq \beta$).
For the $O_h$ symmetry, we introduce
\begin{align}
&O_u = \frac{1}{\sqrt{3}} \left( 3 J_z^2 - \bJ^2 \right),~~~
\varepsilon_u = \frac{1}{\sqrt{3}} \left( 2 \varepsilon_{zz} - \varepsilon_{xx} - \varepsilon_{yy} \right),
\label{eqn:oh-u} \\
&O_v = J_x^2 - J_y^2,~~~~~~~~~~~~~~~~
\varepsilon_v = \varepsilon_{xx} - \varepsilon_{yy}.
\label{eqn:oh-v}
\end{align}
Substituting Eqs. (\ref{eqn:oh-u}) and (\ref{eqn:oh-v}) into Eq. (\ref{eqn:H-oh}), we obtain
\begin{align}
\H_\varepsilon
&= \frac{1}{2} \left( K_{11} - K_{12} \right) \left( \varepsilon_u O_u + \varepsilon_v O_v \right) \cr
&+ \frac{1}{2} K_{44} \left( 2\varepsilon_{yz} O_{yz} + 2\varepsilon_{zx} O_{zx} + 2\varepsilon_{xy} O_{xy}  \right) \cr
&+ \frac{1}{3} \left( K_{11} + 2 K_{12} \right) \left( \varepsilon_{xx} + \varepsilon_{yy} + \varepsilon_{zz} \right) \bJ^2.
\label{eqn:H-oh-2}
\end{align}
Here, the last term proportional to $\bJ^2$ represents a common energy shift.

\subsubsection{$D_{4h}$ symmetry}

The quadrupole--strain coupling for $D_{4h}$ is expressed in the following form:
\cite{Nye-1985,Nowick-1996,Powell-2010}
\begin{align}
\H_\varepsilon
&= K_{11} \left( \varepsilon_{xx} O_{xx} + \varepsilon_{yy} O_{yy} \right) + K_{33} \varepsilon_{zz} O_{zz} \cr
&+ K_{12} \left( \varepsilon_{yy} O_{xx} + \varepsilon_{xx} O_{yy} \right) \cr
&+ K_{31} \varepsilon_{zz} \left( O_{xx} + O_{yy} \right)
  + K_{13} \left( \varepsilon_{xx} + \varepsilon_{yy} \right) O_{zz} \cr
&+ \frac{1}{2} K_{44} \left( 2\varepsilon_{yz} O_{yz} + 2\varepsilon_{zx} O_{zx} \right)
+ \frac{1}{2} K_{66} 2\varepsilon_{xy} O_{xy}.
\label{eqn:H-d4h}
\end{align}
Here, $K_{11}$, $K_{33}$, $K_{12}$, $K_{13}$, $K_{31}$, $K_{44}$, and $K_{66}$ are arbitrary real constants.
Substituting Eq. (\ref{eqn:oh-v}) into Eq. (\ref{eqn:H-d4h}), we obtain
\begin{align}
\H_\varepsilon
&= \frac{1}{2} \left( K_{11} - K_{12} \right) \varepsilon_v O_v \cr
&+ \left( K_{13} - \frac{K_{11} + K_{12}}{2} \right) \left( \varepsilon_{xx} + \varepsilon_{yy} \right) O_{zz}
  + \left( K_{33} - K_{31} \right) \varepsilon_{zz} O_{zz} \cr
&+ \frac{1}{2} K_{44} \left( 2\varepsilon_{yz} O_{yz} + 2\varepsilon_{zx} O_{zx} \right)
  + \frac{1}{2} K_{66} 2\varepsilon_{xy} O_{xy} \cr
&+ \frac{1}{2} \left( K_{11} + K_{12} \right) \left( \varepsilon_{xx} + \varepsilon_{yy} \right) \bJ^2
  + K_{31} \varepsilon_{zz} \bJ^2.
\label{eqn:H-d4h-2}
\end{align}
Here, the last two terms represent a common energy shift.

\subsection{Magnetic field dependence of Hamiltonian}
\label{appendix:H-h}

The magnetic field dependence of the Hamiltonian can be discussed in a similar way to the quadrupole--strain coupling.
In the quadratic order of the magnetic field $\bh=g\mu_{\rm B}\bH=(h_x,h_y,h_z)$,
the field dependence is obtained by the replacement
$\varepsilon_{\alpha\beta}\rightarrow h_\alpha h_\beta$ ($\alpha,\beta=x,y,z)$
in the quadrupole--strain coupling.

\subsubsection{$O_h$ symmetry}

The quadratic field dependence of the Hamiltonian for the $O_h$ symmetry is expressed as
\begin{align}
\H_h^{(2)}
&= \frac{1}{2} \left( K_{11} - K_{12} \right) \left( h_u O_u + h_v O_v \right) \cr
&+ \frac{1}{2} K_{44} \left( 2h_{yz} O_{yz} + 2h_{zx} O_{zx} + 2h_{xy} O_{xy} \right) \cr
&+ \frac{1}{3} \left( K_{11} + 2 K_{12} \right) \bh^2 \bJ^2,
\label{eqn:H-h-oh}
\end{align}
where
\begin{align}
&h_u = \frac{1}{\sqrt{3}} \left( 3 h_z^2 - \bh^2 \right),~~~~~~
h_v = h_x^2 - h_y^2, \cr
&h_{yz} = h_y h_z,~~~~~~
h_{zx} = h_z h_x,~~~~~~
h_{xy} = h_x h_y.
\end{align}

\subsubsection{$D_{4h}$ symmetry}

The quadratic field dependence of the Hamiltonian for the $D_{4h}$ symmetry is expressed as
\begin{align}
\H_h^{(2)}
&= \frac{1}{2} \left( K_{11} - K_{12} \right) h_v O_v \cr
&+ \left( K_{13} - \frac{K_{11}+K_{12}}{2} \right) \left( h_x^2 + h_y^2 \right) O_{zz}
  + \left( K_{33} - K_{31} \right) h_z^2 O_{zz} \cr
&+ \frac{1}{2} K_{44} \left( 2 h_{yz} O_{yz} + 2 h_{zx} O_{zx} \right) + \frac{1}{2} K_{66} 2 h_{xy} O_{xy} \cr
&+ \frac{1}{2} \left( K_{11} + K_{12} \right) \left( h_x^2 + h_y^2 \right) \bJ^2 + K_{31} h_z^2 \bJ^2.
\label{eqn:H-h-d4h}
\end{align}

\section{Weak Coupling Theory for Transition Probability}
\label{appendix:weak}

When the energy splitting in the two-level system is close to $\varepsilon_0 = E_2 - E_1 \simeq n\omega$,
we can focus on the almost degenerate Floquet states $\ket{E_1,0}$ and $\ket{E_2,n}$.
The transition probability is then given by
\cite{Shirley-1965,Aravind-1984,Ho-1985,Son-2009,Koga-2020}
\begin{align}
P_{E_1\rightarrow E_2}^{(n)}(\varepsilon_0,t)
&= \left| \braket{E_2,n|e^{-i\H_2 t} |E_1,0} \right|^2 \cr
&= \frac{v_{-n}^2}{\tilde{q}^2} \sin^2(\tilde{q}t) \cr
&= \frac{v_{-n}^2}{\tilde{q}^2} \frac{1}{2} \left[ 1 - \cos(2\tilde{q}t) \right],
\label{eqn:p-shirley}
\end{align}
with
\begin{align}
&v_{-n}^2
= \frac{1}{4} \left[ n\omega \frac{A_T}{A_L} J_{-n}\left(\frac{A_L}{\omega}\right) \right]^2
= \frac{1}{4} \left[ n\omega \frac{A_T}{A_L} J_{n}\left(\frac{A_L}{\omega}\right) \right]^2, \cr
&\tilde{q} = \sqrt{ v_{-n}^2 + \frac{1}{4}(n\omega - \varepsilon_0 + 2\delta_n)^2}, \cr
&\delta_n = - \sum_{k\ne -n} \frac{v_k^2}{(n+k)\omega}.
\end{align}
Here, $J_n(x)$ represents the Bessel function of the first kind.
In Eq. (\ref{eqn:p-shirley}), $\H_2$ means the Hamiltonian for the two almost degenerate states
expressed in a $2\times 2$ matrix form.
\cite{Shirley-1965}
The numerator $v_{-n}^2$ represents the intensity of the transition.
The present formulation is taken up to the quadratic order of the transverse component $A_T^2$.
The transition probability in Eq. (\ref{eqn:p-shirley}) periodically oscillates in time.

\subsection{Time-averaged transition probability}

Shirley introduced the transition probability as the time-averaged value.
After taking the time-average of Eq. (\ref{eqn:p-shirley}) over a long period,
we can analytically express the time-averaged transition probability for $|A_T|/\omega \ll 1$ as
\cite{Shirley-1965,Son-2009,Koga-2020}
\begin{align}
\bar{P}_{E_1\rightarrow E_2}^{(n)}(\varepsilon_0)
&=\lim_{T\rightarrow \infty} \frac{1}{T} \int_0^T dt P_{E_1\rightarrow E_2}^{(n)}(\varepsilon_0,t) \cr
&= \frac{1}{2}
\frac{v_{-n}^2}{v_{-n}^2 + \frac{1}{4}(n\omega - \varepsilon_0 + 2\delta_n)^2}.
\label{eqn:P-ana}
\end{align}
For $A_L/\omega \ll 1$, we can use the form
\begin{align}
J_n\left(\frac{A_L}{\omega}\right)
\simeq \frac{1}{2^n n!} \left( \frac{A_L}{\omega} \right)^n.
\end{align}
In the weak coupling limit, $v_{-n}^2$ is given by
\begin{align}
&v_0^2 = 0,~~~~~~
v_{-1}^2 = \frac{1}{16} A_T^2, \cr
&v_{-n}^2 = \left[ \frac{A_T}{2^{n+1} (n-1)!} \left( \frac{A_L}{\omega} \right)^{n-1} \right]^2.~~~~~~(n\ge 2)
\label{eqn:vn}
\end{align}
To be precise, third-order ($A_T^3$) terms appear in $v_{-n}^2$ for $n\ge 3$, as reported in Ref. \ref{ref:Gromov-2000},
where $v_{-n}$ and $2\delta_n$ for an electron paramagnetic resonance
were studied by the perturbation theory for $n=2$ and $n=3$.
In our treatment, only terms up to the quadratic order ($A_T^2$) term are taken into account,
and thus the present estimation of $v_{-n}^2$ is valid for $n\le 2$.

Up to the second order of $A/\omega$ in Eq. (\ref{eqn:A}), $\delta_n$ is expressed as
\begin{align}
\delta_1 = - \frac{A_T^2}{32\omega},~~~~~~
\delta_n = - \frac{n}{8(n^2-1)} \frac{A_T^2}{\omega}.~~~~~~(n\ge 2)
\label{eqn:delta}
\end{align}
Substituting Eqs. (\ref{eqn:vn}) and (\ref{eqn:delta}) into Eq. (\ref{eqn:P-ana}),
we obtain the transition probability at $\varepsilon_0=n\omega$ as
\begin{align}
&\bar{P}_{E_1\rightarrow E_2}^{(1)}(\varepsilon_0=\omega) = \frac{1}{2} \frac{1}{1 + \left(\frac{A_T}{8\omega}\right)^2},
~~~~~~~~~~~~~~~~~~~~~~~
(n=1) \cr
&\bar{P}_{E_1\rightarrow E_2}^{(n)}(\varepsilon_0=n\omega)
= \frac{1}{2} \frac{\left(\frac{A_L}{\omega}\right)^{2(n-1)}}
{\left(\frac{A_L}{\omega}\right)^{2(n-1)} + C_n^2\left(\frac{A_T}{\omega}\right)^2},~~~(n\ge 2)
\label{eqn:P-ana-12}
\end{align}
where
\begin{align}
C_n = \frac{2^{n-2} n!}{n^2-1}.
\end{align}
In the weak coupling limit, we obtain the following leading terms:
\begin{align}
&\bar{P}_{E_1\rightarrow E_2}^{(1)}(\varepsilon_0=\omega) \rightarrow \frac{1}{2},
~~~~~~~~~~~~~~~~~~~~~~~~~~~~(n=1) \cr
&\bar{P}_{E_1\rightarrow E_2}^{(2)}(\varepsilon_0=2\omega) = \frac{1}{2} \frac{A_L^2}{A_L^2 + \left(\frac{2}{3}\right)^2 A_T^2},
~~~~~~(n=2)
\label{eqn:analytic} \\
&\bar{P}_{E_1\rightarrow E_2}^{(n)}(\varepsilon_0=n\omega) \rightarrow 0.
~~~~~~~~~~~~~~~~~~~~~~~~~~~~(n\ge 3)
\nonumber
\end{align}

\subsection{Fermi's golden rule}

Let us discuss the relation of the time-averaged transition probability to Fermi's golden rule.
From Eq. (\ref{eqn:p-shirley}), the transition probability for Fermi's golden rule
is defined as the gradient with respect to time as
\begin{align}
&\bar{P}_{\rm Fermi}^{(n)}(\varepsilon_0)
=\lim_{T\rightarrow \infty} \frac{P_{E_1\rightarrow E_2}^{(n)}(\varepsilon_0,T)}{T} \cr
&= v_{-n}^2 \frac{1}{4\left(\frac{\tilde{q}}{2}\right)^2} \lim_{T\rightarrow \infty} \frac{\sin^2(\tilde{q}T)}{T} \cr
&= 2\pi v_{-n}^2 \delta\left( \sqrt{ \left(2v_{-n}\right)^2 + \left(n\omega - \varepsilon_0 + 2\delta_n\right)^2} \right).
\end{align}
Here, $\delta(x)$ represents the Dirac delta function.
In the weak coupling limit ($v_{-n}\rightarrow 0$),
the transition probability takes the following form:
\begin{align}
\bar{P}_{\rm Fermi}^{(n)}(\varepsilon_0) = 2\pi v_{-n}^2 \delta\left( n\omega - \varepsilon_0 + 2\delta_n \right).
\end{align}
This corresponds to Fermi's golden rule.
For the 1-phonon process ($n=1)$, the level shift $2\delta_1$ is known as the Bloch--Siegert shift.
\cite{Bloch-1940}

\subsection{Two-level energy splitting in present model}

For the $S=1$ model studied in Sect. \ref{sec:s=1},
the energy splitting of the two-level system under a magnetic field is
$\varepsilon_0 \rightarrow h^2/D=(g\mu_{\rm B}H)^2/D$.
It is $\varepsilon_0 \rightarrow c(\theta)h^2$ for the $J=4$ model in $O_h$ symmetry studied in Sect. \ref{sec:j=4-oh},
and the Van Vleck shift is anisotropic with respect to the angle $\theta$ of the magnetic field applied in the $xy$-plane.
For the $D_{4h}$ symmetry studied in Sect. \ref{sec:d4h},
it is $\varepsilon_0 \rightarrow \tilde{c}(\theta,\varphi)h^2$
and the Van Vleck shift also depends on the additional parameter $\varphi$
for the wave function of the non-Kramers doublet.
In the case of the quadrupole ordered phase studied in Sect. \ref{sec:order},
it is $\varepsilon_0 \rightarrow h_{\rm eff}^2 + 2\bar{O}_u$.
The quadrupole order parameter $\bar{O}_u$ gives rise to spontaneous energy splitting,
and the resonance frequency $\omega$ depends on $\bar{O}_u$.



\begin{thebibliography}{99}
\bibitem{note:Bloch-Siegert}
For $n=1$ (1-phonon process), this type of the level shift $2\delta_1$ is known as the Bloch--Siegert shift.
\bibitem{note:c}
Here, we assumed a positive coupling constant ($c>0$) in Eq. (\ref{eqn:H-order}).
When $c<0$, the $O_u$ order can be stabilized by a magnetic field $\bh\propto (1,1,0)$.
\bibitem{note:c2}
When $c<0$ in Eq. (\ref{eqn:H-order}),
the $O_v$ order can be stabilized by a magnetic field $\bh\propto (0,\sqrt{2},1)$.
\end{thebibliography}
\end{document}